\documentclass[12pt,a4paper]{article}
\pdfoutput=1

\usepackage{macros}
\usepackage{booktabs}
\usepackage{multirow,makecell}

\preprint{ }
\title{Instantons on Young diagrams with matters}

\author[a]{Yanyan Chen,}
\author[a]{Jiaqun Jiang,}
\author[a]{Satoshi Nawata}
\author[b]{and Yilu Shao}

\affiliation[a]{Department of Physics and Center for Field Theory \& Particle Physics, Fudan University, 20005, Songhu Road, 200438 Shanghai, China}
\affiliation[b]{Institut de Mathématiques de Bourgogne, Université de Bourgogne Franche-Comté,\\9 avenue Alain Savary, Dijon, France
}

\emailAdd{yanyanchen235@gmail.com}
\emailAdd{jiangjiaqun@gmail.com}
\emailAdd{snawata@gmail.com}
\emailAdd{shaoyilu1999@gmail.com}

\abstract{We present the unrefined instanton partition functions of various 5d gauge theories with matter beyond the fundamental representation as sums over Young diagrams. By using these explicit expressions, we verify a range of identities among the instanton partition functions predicted by Higgsing procedures of fivebrane web diagrams and representation theory.}

\begin{document}

\allowdisplaybreaks

\maketitle

\section{Introduction}

One of the ultimate goals of quantum field theory is to obtain exact results, including non-perturbative effects. In recent decades, significant progress has been made in understanding the non-perturbative nature of quantum field theory. Starting with the seminal work by Seiberg and Witten \cite{Seiberg:1994aj,Seiberg:1994rs}, deep insights have been gained into supersymmetric theories with eight supercharges. In particular, Nekrasov performed the first instance of supersymmetric localization for instanton partition functions \cite{Nekrasov:2002qd}, providing exact results for certain observables in the Seiberg-Witten theory, such as the low-energy effective prepotential.

More remarkably, the fixed points of equivariant actions on the instanton moduli spaces are classified by a set of Young diagrams, and the partition function is expressed as a sum over Young diagrams. Nekrasov's exact result has far-reaching consequences in both physics and mathematics. For example, the expression as a sum over Young diagrams is important because it provides a connection to the topological vertex \cite{AKMV,Awata:2005fa,IKV}. It is also particularly useful in the context of the AGT relation \cite{Alday:2009aq}, where the instanton partition function is identified with conformal blocks of a 2d CFT.

These developments were exclusively made in the gauge groups of $A$ type until recent years. However, closed-form expressions of the unrefined instanton partition functions for pure Yang-Mills theory with gauge groups of $B C D$ type were recently obtained as sums over Young diagrams \cite{Nawata:2021dlk}. Additionally, the connection to the topological vertex \cite{Hayashi:2020hhb} of an O5-plane was uncovered.

In this paper, we push forward the research direction in \cite{Nawata:2021dlk}, aiming to obtain unrefined instanton partition functions with a hypermultiplet beyond the fundamental representation as sums over 2d and 4d Young diagrams.
We combine two approaches to studying 5d supersymmetric gauge theories with matters beyond the fundamental representation. The first approach is the ADHM description, which has been studied in previous works such as \cite{Nekrasov-Shadchin, Shadchin:2005mx, Kim:2012gu, Kim:2012qf, Hwang:2014uwa}. The second approach is the fivebrane construction, which has been extensively studied in \cite{Bergman:2014kza, Bergman:2015dpa, Zafrir:2015ftn, Hayashi:2016jak, Kim-Yagi, Kim:2018gjo, Hayashi:2018bkd, Hayashi:2018lyv, Cheng:2018wll, Hayashi:2019yxj,Kim:2021cua,Kim:2022dbr}. Our main results are as follows: first, we demonstrate that the poles of the JK residue integral arising from the ADHM descriptions can be classified by 2d and 4d Young diagrams in the unrefined limit. Second, we verify identities of the instanton partition functions predicted by Higgsing procedures of fivebrane web diagrams and representation theory.

The paper is structured as follows. In  \S\ref{sec:SU-sym},  we investigate the $\SU(N)$ gauge theory with (anti-)symmetric hypermultiplet, and we represent the unrefined instanton partition functions as sums over 2d Young diagrams. In  \S\ref{sec:SO-spinor}, we study the $\SO(N)$ gauge theory with (conjugate) spinor hypermultiplet and similarly express the unrefined instanton partition functions as sums over 2d Young diagrams. In \S\ref{sec:SO-max}, we consider the $\cN=1^*$ $\SO(N)$ gauge theory, and in \S\ref{sec:Sp-anti}, we investigate the $\Sp(N)$ gauge theory with antisymmetric hypermultiplet. In these latter cases, the unrefined instanton partition functions are represented as sums over 4d Young diagrams. In all of these cases, we verify the identities of the instanton partition functions predicted by Higgsing of fivebrane webs and representation theory.

\section{\texorpdfstring{$\SU(N)$}{SU(N)} gauge group with (anti-)symmetric hypermultiplet}\label{sec:SU-sym}

The instanton partition function can be obtained from the equivariant Chern characters of the universal bundle $\cE$ over the instanton moduli space \cite{Nekrasov:2002qd}.
Let $p$ be a generic element of the equivariant torus action on the universal bundle $\cE$. The equivariant Chern characters for (rank-two) symmetric and antisymmetric representations of $\SU(N)$ can be obtained by
$$
\begin{aligned}
\mathrm{Ch}_p^{\text {sym }}(\mathcal{E}) &=\mathrm{Ch}_p\left(\mathrm{Sym}^2 \mathcal{E}\right)=\frac{1}{2}\left[(\mathrm{Ch}_p(\mathcal{E}))^2 + \mathrm{Ch}_{p^2}(\mathcal{E})\right], \\
\mathrm{Ch}_p^{\text {ant }}(\mathcal{E}) &=\mathrm{Ch}_p\left(\wedge^2 \mathcal{E}\right)=\frac{1}{2}\left[(\mathrm{Ch}_p(\mathcal{E}))^2 - \mathrm{Ch}_{p^2}(\mathcal{E})\right]~.
\end{aligned}
$$
The equivariant index of the Dirac operator on the universal bundle takes the following form
$$
\operatorname{Ind}_q=\sum_\alpha \epsilon_\alpha \mathrm{e}^{w_a}=\int_{\mathbb{C}^2} \operatorname{Ch}_q(\mathcal{E}) \mathrm{Td}_q\left(\mathbb{C}^2\right)_t~.
$$
Then, the contributions to the integrands of the 5d instanton partition function can be read off by mapping
$$
\sum_\alpha \epsilon_\alpha \mathrm{e}^{w_\alpha} \mapsto \prod_\alpha \sh^{\epsilon_\alpha} (w_\alpha) ~.
$$
We refer the reader to \cite{Nekrasov:2002qd,Losev:2003py,Shadchin:2005mx} for the details. Note that we use the notation that
\begin{equation}
    \sh(x):=\mathrm{e}^{\frac{x}{2}}-\mathrm{e}^{-\frac{x}{2}}~,\qquad \ch(x):=\mathrm{e}^{\frac{x}{2}}+\mathrm{e}^{-\frac{x}{2}}~.
\end{equation}

In this way, we can write down the contour integral expressions of the instanton partition functions
\be
Z_{\SU(N),k,\kappa}^{\textrm{rep}}=\frac{1}{k ! 2^k} \oint_{\textrm{JK}} \prod_{I=1}^k \frac{d \phi_I}{2 \pi i} \cdot e^{\kappa \sum_{I=1}^k \phi_I}\cI_{\SU(N),k}^{\textrm{vec}}\cI_{\SU(N),k}^{\textrm{rep}}
\ee
where the integrands are given by
\begin{align}
\cI_{\SU(N),k}^{\textrm{vec}}=& \frac{\prod_{I \neq J} \sh( \phi_{I}-\phi_{J}) \cdot \prod_{I, J} \sh (2 \epsilon_{+}-\phi_{I}+\phi_{J})}{\prod_{I, J} \sh (\epsilon_{1,2}+\phi_{I}-\phi_{J}) \prod_{I=1}^{k}  \prod_{s=1}^{N} \sh (\epsilon_{+} \pm( \phi_{I} - a_{s}))}\cr
\cI_{\SU(N),k}^{\textrm{fund}}(m)=&\prod_{I=1}^k\sh(\phi_I+ m) \cr
\cI_{\SU(N),k}^{\textrm{sym}}(m)=& \prod_{I=1}^{k} \sh{(2\phi_I+ m \pm \e_-)} \prod_{s=1}^{N} \sh{(\phi_I + a_s + m)} \prod_{I<J}^{k} \frac{\sh{(\phi_I + \phi_J +m\pm \e_-)}}{\sh{(-\epsilon_+ \pm(\phi_I + \phi_J+ m)) }}\cr
\cI_{\SU(N),k}^{\textrm{anti}}(m)=& \prod_{I=1}^{k}\frac{ \prod_{s=1}^{N} \sh{(\phi_I + a_s + m)}}{ \sh{(-\epsilon_+\pm(2\phi_I+ m))}} \prod_{I<J}^{k} \frac{\sh{(\phi_I + \phi_J +m\pm \e_-)}}{\sh{(-\epsilon_+ \pm(\phi_I + \phi_J+ m))}}~.
\end{align}
Note that $\kappa$ is the 5d Chern-Simons level. For a gauge group $\SU(2N+1)$ of even rank, the level $\kappa$ needs to be a half-odd-integer due to the parity anomaly associated with the (anti-)symmetric matter.
The residue computation is carried out by the Jeffrey-Kirwan prescription \cite{jeffrey1995localization,Benini:2013xpa}. With (anti-)symmetric hypermultiplet, it is difficult to classify poles fully and obtain closed-form expressions at the refined level. Nonetheless, one can check that the following isomorphisms of representations of gauge Lie algebras\footnote{We perform the following change of variables on the Coulomb branch parameters of the instanton partition function from  the orthogonal basis of $\SU(4)$  to that  of $\SO(6)$:
$$    A_1\to(A_1A_2A_3)^{\frac{1}{2}},\quad A_2\to(A_1A_2^{-1}A_3^{-1})^{\frac{1}{2}},\quad A_3\to(A_1^{-1}A_2A_3^{-1})^{\frac{1}{2}},\quad A_4\to(A_1^{-1}A_2^{-1}A_3)^{\frac{1}{2}}~.$$
}
\begin{align}
(\fraksu(4),\AS)\cong (\frakso(6),\mathbf{V}) \quad \longrightarrow \quad
& Z_{\SU(4),k,\kappa=0}^{\textrm{anti}}=Z_{\SO(6),k}^{\textrm{vect}} ~,\\
(\fraksu(3),\AS)\cong (\fraksu(3),\mathbf{F}) \quad \longrightarrow\quad
& Z_{\SU(3),k,\kappa=\frac12}^{\textrm{anti}}(m)=Z_{\SU(3),k,\kappa=\frac12}^{\textrm{fund}} (m\to-m)~.\nonumber
\end{align}
We verify these identities up to 4-instanton by performing JK residues explicitly at the refined level.

\begin{figure}[ht]
    \centering
   \includegraphics[width=6.2cm]{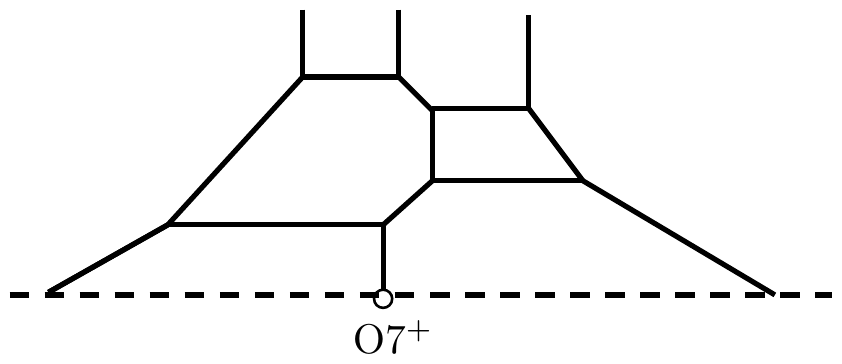}\quad     \includegraphics[width=8.2cm]{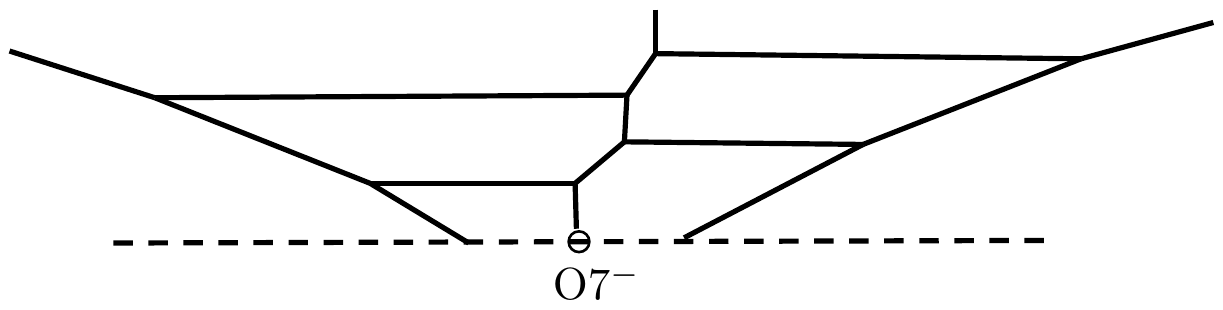}
    \caption{(Left-panel) $\SU(4)$ gauge theory with a symmetric hypermultiplet. \\ (Right-panel) $\SU(4)$ gauge theory with an anti-symmetric hypermultiplet. }
    \label{fig:SUN-anti-sym}
\end{figure}

In \cite{Bergman:2015dpa}, 5d $\SU(N)$ gauge theory with one (anti-)symmetric hypermultiplet is constructed from fivebranes with an O7-plane in Type IIB theory. (See Figure \ref{fig:SUN-anti-sym}.)
Once we align D5-branes in the brane configuration with an O7${}^+$-plane (symmetric hypermultiplet), we can higgs the middle NS5-brane, yielding the brane configuration for pure $\SO(N)$ gauge theory. This operation can be seen at the level of the instanton partition functions as
\be\label{SU-SO}
Z_{\SU(N),2k,\kappa=\frac12 (N \bmod 2)}^{\textrm{sym}}(a_{N-i+1}=-a_i,m=\e_+)=(-1)^{k (N+ 1)}Z_{\SO(N),k}^{\textrm{pure}}~.
\ee
Note that $2k$-instanton on the left-hand side coincides with $k$-instanton on the right-hand side. Also, when $N$ is odd, we set $a_{\frac{N+1}{2}}=0$ in the left-hand side.

Similarly, we can perform the same operation on the brane configuration with an O7${}^-$-plane (anti-symmetric hypermultiplet), which results in  the brane configuration for pure $\Sp(N)$ gauge theory. This operation amounts to the identities of the instanton partition functions
\begin{equation}\label{SU-Sp}
\begin{aligned}
Z_{\SU(2N),k,\kappa=N\bmod 2}^{\textrm{anti}}(a_{N-i+1}=&-a_i,m=\e_+)=(-1)^{k(N+1)+\lceil\frac{k}{2}\rceil}Z_{\Sp(N),k,\theta=0}^{\textrm{pure}}~,\cr
Z_{\SU(2N),k,\kappa=N+1 \bmod 2}^{\textrm{anti}}(a_{N-i+1}=&-a_i,m=\e_+)=(-1)^{k(N+1)+\lceil\frac{k}{2}\rceil}Z_{\Sp(N),k,\theta=\pi }^{\textrm{pure}}~,
\end{aligned}
\end{equation}
where the discrete $\theta$-angle of the $\Sp(N)$ theory is determined by the Chern-Simons level and the rank of the gauge group.

Although pole classification is hard at the refined level, non-trivial JK-poles are classified by $N$-tuples $\vec{\lambda}=(\lambda^{(1)},\ldots,\lambda^{(N)})$ of 2d Young diagrams with the total number of boxes $\sum_{s=1}^N|\lambda^{(s)}|=k$ in the \emph{unrefined} limit $\e_1=-\e_2=\hbar$.
We specify a pole location associated with a content $x=(x_1,x_2)\in \lambda^{(s)}$ as
\begin{equation}\label{phi(s)}
  \phi_{s}(x)= a_s+(x_1-x_2)\hbar\  .
\end{equation}
By writing the Nekrasov factor \cite{Nekrasov:2002qd}
\begin{equation}\label{Nij}
  N_{s,t}(x):= a_s- a_t-\hbar(a_{\lambda^{(s)}}(x)+l_{\lambda^{(t)}}(x)+1)\ ,
\end{equation}
the residue sum for a symmetric hypermultiplet is expressed as
\begin{equation}\label{SU-sym}
\begin{aligned}
&Z_{\SU(N),k,\kappa}^{\textrm{sym}}\cr
=&(-1)^{kN+\lceil\frac{k}{2}\rceil}\sum_{\vec{\lambda}} \prod_{s=1}^{N} \prod_{x \in \lambda^{(s)}} e^{\kappa \phi_{s}(x)}\frac{\sh(2\phi_{s}(x)+m\pm \hbar)\cdot  \prod_{t=1}^{N}\sh(\phi_{s}(x)+a_t+m)}{\prod_{t=1}^{N} \sh ^{2}(N_{s,t}(x))}\\
&\times \prod_{s \leq t}^{N}
\prod_{\substack{x \in \lambda^{(s)},y \in \lambda^{(t)} \\x<y}}
\frac{\sh(\phi_{s}(x)+\phi_{t}(y)+m\pm \hbar)}{\sh ^{2}(\phi_{s}(x)+\phi_{t}(y)+m)}
\end{aligned}
\end{equation}
Here we define a total ordering on the boxes of 2d Young diagrams:
\be\label{total-ordering}
\lambda^{(s)} \ni x<y \in \lambda^{(t)} \quad \text { if }\left\{\begin{array}{l}
s<t \\
s=t, x_1<y_1 \\
s=t, x_1=y_1 , x_2<y_2
\end{array}\right.~.
\ee
Once we take the specialization \eqref{SU-SO}  at the unrefined level, we obtain an expression as a Young diagram sum for the pure $\SO(N)$ instanton partition function, which is different from the one in \cite{Nawata:2021dlk}.

For an anti-symmetric matter,  the JK-poles are classified by $(N+4)$-tuples of 2d Young diagrams with the total number of boxes $\sum_{s=1}^{N+4}|\lambda^{(s)}|=k$ where the four additional effective Coulomb branch parameters are
\be \label{additional-Coulomb}
a_{N+j}= -\frac{m}{2} (+\pi i )~, \   \frac{\hbar-m}{2} (+\pi i )~,
\ee
for $j=1,2,3,4$.
Then, the residue sums are expressed as
\begin{align}\label{SU-anti}
&Z_{\SU(N),k,\kappa}^{\textrm{anti}}(A_{N+1}=e^{\frac{m}{2}},A_{N+2}=-e^{\frac{m}{2}},A_{N+3}=e^{-\frac{\hbar-m}{2}},A_{N+4}=-e^{-\frac{\hbar-m}{2}})\cr
=&(-1)^{kN+\lceil\frac{k}{2}\rceil}\sum_{\vec{\lambda}} C^{\textrm{anti}}_{\vec{\lambda}, \vec{A}}  \prod_{s=1}^{N+4} \prod_{x \in \lambda^{(s)}}e^{\kappa \phi_{s}(x)}
\frac{\sh^2(2\phi_{s}(x)+m-\hbar)\cdot\prod_{t=1}^{N}\sh(\phi_{s}(x)+a_t+m)}{\prod_{t=1}^{N+4} \sh ^{2}(N_{s,t}(x))}\cr
&\qquad\qquad\qquad \times \prod_{s \leq t}^{N+4}
\prod_{\substack{x \in \lambda^{(s)},y \in \lambda^{(t)} \\x<y}}
\frac{\sh(\phi_{s}(x)+\phi_{t}(y)+m\pm\hbar)}{\sh^{2}(\phi_{s}(x)+\phi_{t}(y)+m)}
\end{align}
where $\phi_{s}(x)$ and $N_{s,t}(x)$ are given in \eqref{phi(s)} and \eqref{Nij}. Here, the non-trivial multiplicity constants
\be\label{Dweight}
C^{\textrm{anti}}_{\vec{\lambda},\vec{A}}=\prod_{s=N+1}^{N+4}C^{\textrm{anti}}_{\lambda^{(s)},A_s}~.
\ee
are involved due to higher-order poles coming from the additional effective Coulomb branch parameters \eqref{additional-Coulomb}. These multiplicity constants are in principle determined by the orders of poles and the determinant of JK vectors that define a cone. Here we conjecture these constants as follows: $C^{\textrm{anti}}_{\lambda^{(s)}=\emptyset,A_s}=1$, and
\be\label{C-anti}
\begin{split}
{\raisebox{-1.5cm}{\includegraphics[width=5cm]{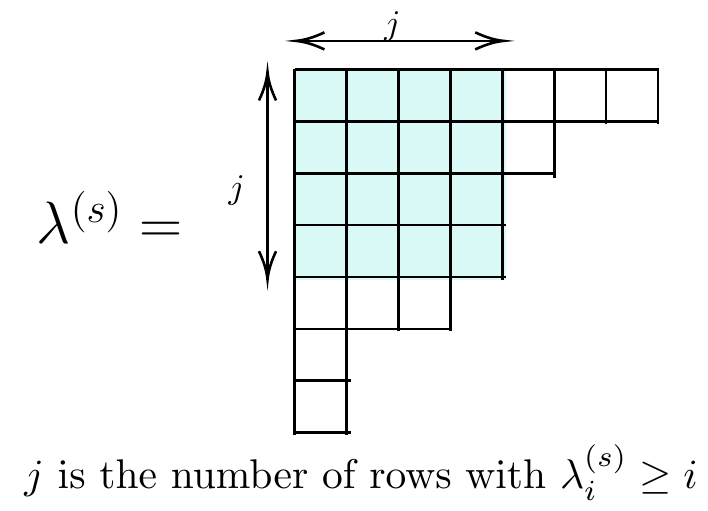}}} \quad \leadsto \quad C^{\textrm{anti}}_{\lambda^{(s)},A_s=\pm e^{-\frac{m}2}}=(-2)^{j}\\
{\raisebox{-1.5cm}{\includegraphics[width=5cm]{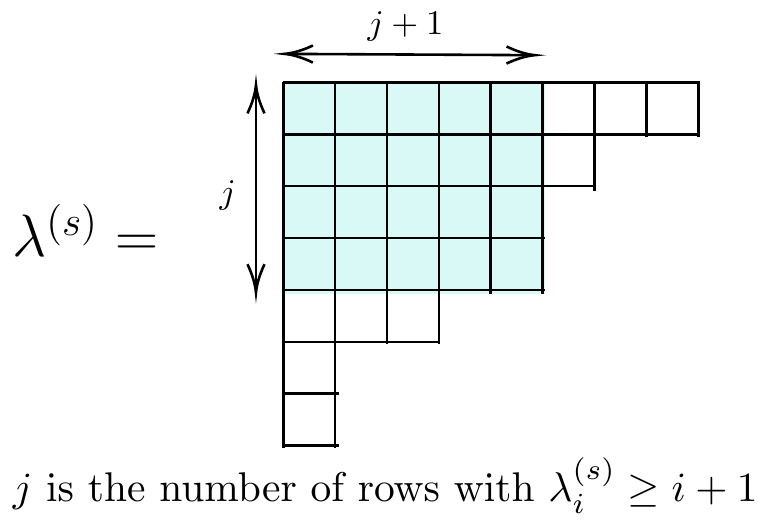}}} \quad \leadsto \quad C^{\textrm{anti}}_{\lambda^{(s)},A_s=\pm e^{-\frac{\hbar-m}2}}=(-2)^{j}
\end{split}\ee
Once we take the specialization \eqref{SU-Sp}  at the unrefined level, we obtain an expression as Young diagram sums for the pure $\Sp(N)$ instanton partition function (both $\theta=0$ and $\theta=\pi$), which takes a different presentation from the one in \cite{Nawata:2021dlk}. 

Since the relation to the instanton partition functions for gauge groups of $BCD$ comes from the fivebrane webs with an O7-plane here, the difference in the expressions of the partition functions from \cite{Nawata:2021dlk} may be due to the distinction between O7-planes and O5-planes in the topological vertex formalism. Further investigation from this perspective could be worthwhile.

The relationship between the instanton partition functions for the symmetric and antisymmetric hypermultiplet, as viewed from the perspective of fivebrane systems with an O-plane, will be explored in the upcoming work \cite{Chengdu}. This will also help to deepen our understanding of the additional Coulomb branch parameters \eqref{additional-Coulomb} and the multiplicity coefficients \eqref{C-anti}.

\section{\texorpdfstring{$\SO(n)$}{SO(n)} gauge group with spinor hypermultiplet}\label{sec:SO-spinor}

\begin{figure}[ht]
    \centering
   \includegraphics[width=6.2cm]{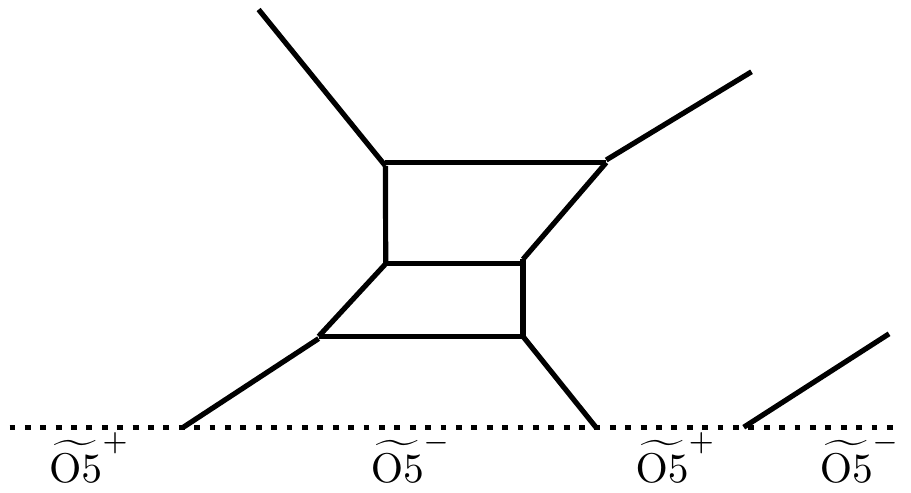}\quad     \includegraphics[width=8.2cm]{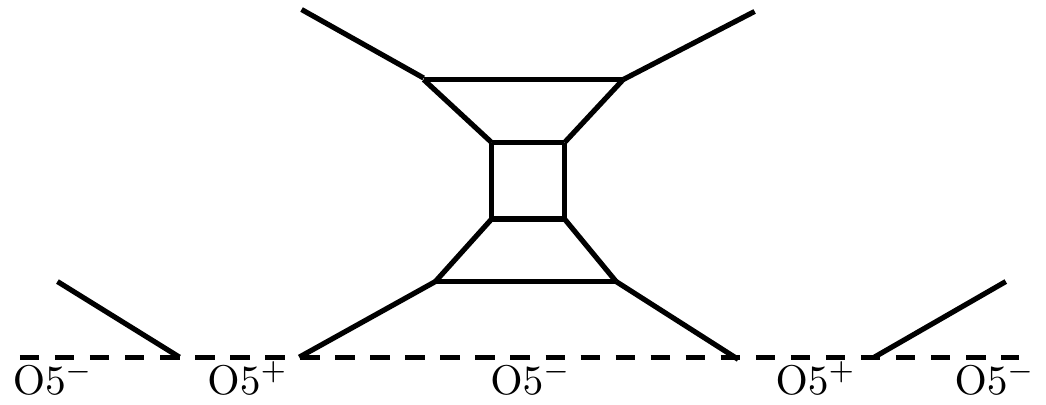}
    \caption{(Left-panel) $\SO(7)$ gauge theory with a spinor matter. \\ (Right-panel) $\SO(8)$ gauge theory with spinor and conjugate spinor matters. }
    \label{fig:SO-spinor}
\end{figure}

  A Type IIB brane configuration for a hypermultiplet in the spinor representation of a 5d SO($n$) gauge theory is given in \cite{Zafrir:2015ftn} by using O5-planes.  To introduce two hypermultiplets in the spinor representation, a ``leg'' fivebrane needs to be introduced on the right side. For instance, we draw the fivebrane diagrams for SO(7) gauge group with one spinor, and SO(8) gauge theory with spinor and conjugate spinor  in Figure \ref{fig:SO-spinor}. As the figures show, the diagram for a hypermultiplet in the (conjugate) spinor representation  can be intuitively interpreted as the “Sp(0)” gauge theory. In other words, roughly speaking, the theory can be regarded as an “SO($n$)-Sp(0) quiver” gauge theory.

\begin{figure}[ht]
    \centering
   \includegraphics[width=\textwidth]{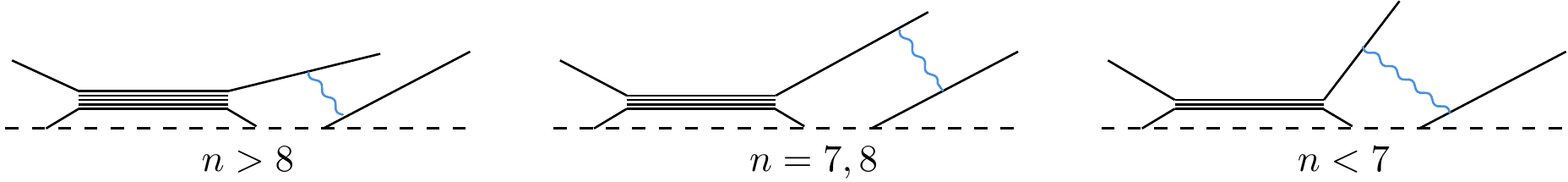}
    \caption{D1'-branes (blue) create hypermultiplet particles in the spinor representation.} 
    \label{fig:legs}
\end{figure}

Instantons are created by D1-branes that are suspended by fivebranes in the middle. On the other hand, hypermultiplet particles in the spinor representation are created by D1'-branes that are suspended by ``leg'' fivebranes.
As illustrated in Figure \ref{fig:legs}, for SO($n$) gauge theory with $n \leq 6$, the ``leg'' fivebranes do not intersect. For $n=7,8$, the two ``leg'' fivebranes are parallel to each other.
In fact, from this viewpoint of quantum mechanics on D1-D1'-branes, the SO($n$) instanton moduli spaces with spinor matter are described using the ADHM approach in \cite{Kim:2018gjo}. The results are reviewed in Appendix \ref{app:integral}.

When $k$ D1-branes create $k$ instantons and $j$ D1'-branes create $j$ hypermultiplet particles, we can view the system as $\Sp(k)$-$\OO(j)$ quiver gauge theory from the perspective of quantum mechanics on the D1-D1'-branes. (See Figures \ref{fig:ADHM-odd} and \ref{fig:ADHM-even}.) The gauge group O($j$) has two connected components, $\OO(j)_\pm$, which leads to generally two contributions, $Z^{\SO(n)}_{k,j,\pm}$, to the partition function for a (conjugate) spinor matter. These contributions can be obtained by performing JK residue integrals, as detailed in Appendix \ref{app:integral}.
However, for $n=\textrm{odd}$, $Z^{\SO(n)}_{k,\textrm{even},-}$ and $Z^{\SO(n)}_{k,\textrm{odd},+}$ vanish due to fermionic zero modes. The full partition function for a spinor hypermultiplet can still be written as
\begin{equation}\label{spinor}
Z_{\textrm{full},\SO(n)}^{\textrm{spinor}}=\sum_{k,j=0}^{\infty}\frakq^k e^{-jm}\frac{Z^{\SO(n)}_{k,j,+}+ Z^{\SO(n)}_{k,j,-}}{2}
\end{equation}
where $m$ is the mass of the hypermultiplet.
On the other hand, for a conjugate spinor hypermultiplet, the full partition function is
\begin{equation}\label{conj}
Z_{\textrm{full},\SO(n)}^{\textrm{conj}}=\sum_{k,j=0}^{\infty}\frakq^k e^{-jm} (-1)^{j}\frac{Z^{\SO(n)}_{k,j,+}+ (-1)^{\mathrm{sign}(j)} Z^{\SO(n)}_{k,j,-}}{2}
\end{equation}
where $\mathrm{sign}(j)=0$ for $j=0$, and $\mathrm{sign}(j)=1$ for $j>0$. For $n<9$, the full partition function $Z_{\textrm{full}}$ contains not only instanton contributions, but also spurious contributions.

As analyzed in \cite{Kim:2018gjo}, to extract genuine instanton contributions, we must remove spurious contributions. Below, we focus only on the spinor hypermultiplet, but the story is parallel for the conjugate spinor (just insert appropriate signs as in \eqref{conj}). 
For $n\leq 6$, the partition function at $k=0$ is given by
  \begin{align}\label{spurious}
  \sum_{j=0}^{\infty}e^{-jm}\frac{Z^{\SO(n)}_{0,j,+}+ Z^{\SO(n)}_{0,j,-}}{2}=Z_{\textrm{pert}}\equiv
  \textrm{PE}\left[e^{-m}\frac{Z^{\SO(n)}_{0,1,+}+ Z^{\SO(n)}_{0,1,-}}{2}\right]
  \end{align}
where PE is the plethystic exponent \eqref{plethystic}. Consequently, the full partition function factorizes as
  \begin{align}
   Z_{\textrm{full},\SO(n)}^{\textrm{spinor}}=\widehat Z_{\SO(n)}^{\textrm{spinor}}(\frakq, \e_{1,2}, m) Z_{\textrm{pert}}(\e_{1,2}, m)
\end{align}
where we extract the genuine instanton contribution
\begin{equation}\label{Zhat}
   \widehat Z_{\SO(n)}^{\textrm{spinor}}(\frakq, \e_{1,2}, m) = \sum_{k=0}^{\infty} \frakq^{k} \widehat Z_{\SO(n),k}^{\textrm{spinor}}(\e_{1,2}, m)~,\quad \widehat Z_{\SO(n),k}^{\textrm{spinor}}(\e_{1,2}, m)=\sum_{j=0}^{\infty}e^{-j m} \widehat Z_{\SO(n),k, j}^{\textrm{spinor}}~.
\end{equation}
For the cases of $n\leq 6$, the contributions from hypermultiplet particles more than instanton numbers vanish as $\widehat{Z}_{k, j}=0$ for $j>k$ for both the spinor and the conjugate spinor.

For $n=7,8$, the partition function at $k=0$ is given by
  \begin{align}\label{spurious2}
  \sum_{j=0}^{\infty}e^{-jm}\frac{Z^{\SO(n)}_{0,j,+}+ Z^{\SO(n)}_{0,j,-}}{2}=&Z_{\textrm{pert}}Z_{\textrm{extra}}\\ =&  \textrm{PE}\left[e^{-m}\frac{Z^{\SO(n)}_{0,1,+}+ Z^{\SO(n)}_{0,1,-}}{2}\right] \mathrm{PE}\left[-\frac{1}{2} e^{-2 m} \frac{ \ch 2\epsilon_{+}}{ \sh\epsilon_{1,2}}\right]~,\nonumber
  \end{align}
  where $Z_{\textrm{extra}}$ is the contribution from D1'-branes running away along the two parallel fivebranes. (See the middle of Figure \ref{fig:legs}.)
Thus, the full partition function factorizes as
\begin{equation}
 Z_{\textrm{full},\SO(n)}^{\textrm{spinor}}=\widehat Z_{\SO(n)}^{\textrm{spinor}}(\frakq, \e_{1,2}, m) Z_{\textrm{pert}}( \e_{1,2}, m) Z_{\text {extra }}( \e_{1,2}, m)~,
\end{equation}
  where $\widehat Z_{\SO(n)}^{\textrm{spinor}}(\frakq, \e_{1,2}, m) $ takes the form \eqref{Zhat}. In the cases of $n=7,8$, for both the spinor and the conjugate spinor, the instanton partition functions are subject to $\widehat{Z}_{k, 2 k-n}=\widehat{Z}_{k, n}$ so that  $\widehat{Z}_{k, j}=0$ for $j>2k$.

For $n\geq 9$, we cannot express the contribution purely from the hypermultiplet in the form of a plethystic exponent, unlike in previous examples \eqref{spurious} and \eqref{spurious2}. This is because the leg fivebranes meet at a certain point in this case. Such a fivebrane configuration has been studied in \cite{Hayashi:2019yxj}. However, we believe that the full partition functions \eqref{spinor} and \eqref{conj} can still be applied to these fivebrane web configurations even for $n\geq 9$, although we leave a more detailed investigation of this issue for future work.

As we learned in \cite{Nawata:2021dlk}, the instanton partition function of pure Yang-Mills theory for gauge groups of $BCD$ type can be written as sums over 2d Young diagrams at the unrefined level. Since the SO($n$) theory with a spinor matter can be roughly regarded as an ``SO($n$)-Sp(0) quiver'' gauge theory, we expect that its instanton partition function can also be expressed as a sum over 2d Young diagrams in the unrefined limit. In the following, we will demonstrate that this is indeed the case. Moreover, we can generalize the method in this section to the SO-Sp linear quiver gauge theories.

\subsection{\texorpdfstring{$\SO(2N+1)$}{SO(2N+1)} with one spinor}

The integral expressions for the partition functions are given in Appendix \ref{app:integral1}.
For a gauge group of $B$ type, the contributions from $\OO(2\ell)_-$ and $\OO(2\ell+1)_{+}$ sectors vanish due to fermionic zero modes.
As a result, we have a single contribution, $Z^{\SO(2N+1)}_{k,j}$, for each hypermultiplet particle number $j$. This is another way to see that there is no distinction between the spinor representation and its conjugate for $\SO(n)$ with odd $n$.

As in \cite{Nawata:2021dlk} and we also discussed below \eqref{SU-anti}, there are four effective Coulomb branch parameters so that even the “Sp(0)” gauge theory contributes to the unrefined partition function. Hence, for $Z^{\SO(2N+1)}_{k,2\ell(+1)}$, the non-trivial JK poles in the unrefined limit are classified $(N+4)$-tuples $\vec\lambda$ of 2d Young diagrams where the numbers of boxes are subject to
\begin{equation}
    \sum_{s=1}^N |\lambda^{(s)}| = k~,\qquad \sum_{s=N+1}^{N+4} |\lambda^{(s)}| = \ell~.
\end{equation}
The JK residue sums are expressed as
 \begin{footnotesize}
     \begin{align}
         \label{SO7Sp-1}
         &Z^{\SO(2N+1)}_{k,2\ell}(A_{N+1}=q^{\frac12},A_{N+2}=-q^{\frac12},A_{N+3}=1,A_{N+4}=-1;q)=(-1)^{k+\ell}
         \cr
         &\sum_{\vec{\lambda}}C^{\Sp}_{\vec{\lambda},\vec{A}}\prod_{s=1}^{N}\prod_{x\in \lambda^{(s)}}
         \frac{\sh^{4}2\phi_{s}(x)}
         { \sh^2(\phi_{s}(x))\prod\limits_{t=1}^{N} \sh^2(N_{s,t}(x))\, \sh^2(\phi_{s}(x)+ a_t)} \prod_{s\leq
         t}^{N}\prod_{\substack{x\in \lambda^{(s)}\\y\in \lambda^{(t)}\\ x< y }}\frac{\sh^4(\phi_{s}(x)+\phi_{t}(y))}
         { \sh^2(\phi_{s}(x)+\phi_{t}(y)\pm\hbar)}
         \cr
         & \prod_{s=N+1}^{N+4}\prod_{x\in \lambda^{(s)}}
         \frac{\sh^2(\phi_{s}(x))\,\sh^{4}(2\phi_{s}(x))\prod\limits_{t=1}^{N}\sh(\pm\phi_{s}(x)+a_t)}
         {\prod\limits_{t=N+1}^{N+4} \sh^2(N_{s,t}(x))\, \sh^2(\phi_{s}(x)+ a_t)} \prod_{N+1\leq s\leq
         t}^{N+4}\prod_{\substack{x\in \lambda^{(s)}\\y\in \lambda^{(t)}\\ x< y }}\frac{\sh^4(\phi_{s}(x)+\phi_{t}(y))}
         { \sh^2(\phi_{s}(x)+\phi_{t}(y)\pm\hbar)}
         \cr
         &\prod_{s=1}^{N}\prod_{x\in \lambda^{(s)}} \prod_{t=N+1}^{N+4}\prod_{y\in
         \lambda^{(t)}}\frac{\sh(\hbar\pm\phi_{s}(x)\pm\phi_{t}(y))}
         {\sh(\pm\phi_{s}(x)\pm\phi_{t}(y))}
     \end{align}
 \end{footnotesize}
 \begin{footnotesize}
     \begin{align}
         \label{SO7Sp-2}
         &Z^{\SO(2N+1)}_{k,2\ell+1}(A_{N+1}=q^{\frac12},A_{N+2}=-q^{\frac12},A_{N+3}=1,A_{N+4}=-q;q)= (-1)^{k+\ell+1}\frac{\prod_{t=1}^3\ch a_t}{\sh^2\hbar}
         \cr
         &\sum_{\vec{\lambda}}C^{\Sp}_{\vec{\lambda},\vec{A}}\prod_{s=1}^{N}\prod_{x\in \lambda^{(s)}}
         \frac{\sh^{4}2\phi_{s}(x)}
         { \sh^2(\phi_{s}(x))\prod\limits_{t=1}^{N} \sh^2(N_{s,t}(x))\,\sh^2(\phi_{s}(x)+ a_t)} \prod_{s\leq
         t}^{N}\prod_{\substack{x\in \lambda^{(s)}\\y\in \lambda^{(t)}\\ x< y }}\frac{\sh^4(\phi_{s}(x)+\phi_{t}(y))}
         { \sh^2(\phi_{s}(x)+\phi_{t}(y)\pm\hbar)}
         \cr
         &\prod_{s=N+1}^{N+4}\prod_{x\in \lambda^{(s)}}
         \frac{\sh^2(\phi_{s}(x))\,\sh^{4}(2\phi_{s}(x))\prod\limits_{t=1}^{N}\sh(\pm\phi_{s}(x)+a_t)}
         {\prod\limits_{t=N+1}^{N+4} \sh^2(N_{s,t}(x))\, \sh^2(\phi_{s}(x)+ a_t)} \prod_{N+1\leq s\leq
         t}^{N+4}\prod_{\substack{x\in \lambda^{(s)}\\y\in \lambda^{(t)}\\ x< y }}\frac{\sh^4(\phi_{s}(x)+\phi_{t}(y))}
         { \sh^2(\phi_{s}(x)+\phi_{t}(y)\pm\hbar)}
         \cr
         &\prod_{s=1}^{N}\prod_{x\in \lambda^{(s)}} \prod_{t=N+1}^{N+4}\prod_{y\in \lambda^{(t)}}\frac{\sh(\hbar\pm\phi_{s}(x)\pm\phi_{t}(y))}
         {\sh(\pm\phi_{s}(x)\pm\phi_{t}(y))}\frac{\ch(\hbar\pm\phi_{s}(x))}{\ch(\pm\phi_{s}(x))}
     \end{align}
 \end{footnotesize}
 where $\phi_{s}(x)$ and $N_{s,t}(x)$ are given in \eqref{phi(s)} and \eqref{Nij}, and $C^{\Sp}_{\vec{\lambda},\vec{A}}$ is a constant
 \be\label{weight}
 C^{\Sp}_{\vec{\lambda},\vec{A}}=\prod_{s=N+1}^{N+4}C^{\Sp}_{\lambda^{(s)},A_s}~.
 \ee
These coefficients are the same as those that appear in the pure $\Sp(N)$ instanton partition function in \cite{Nawata:2021dlk}: $C_{\lambda^{(s)}=\emptyset,A_s}=1$, and
\be\begin{split}\label{weight2}
C^{\Sp}_{\lambda^{(s)},A_s=\pm1,\pm q^{\frac12}}&=\frac{2^{2j-1}}{\binom{2j-1}{j-1}}\qquad  \textrm{  where $j$ is the number of rows with $\lambda_i^{(s)}\ge i$}~,\\
C^{\Sp}_{\lambda^{(s)},A_s=\pm q}&=\frac{2^{2j}}{\binom{2j+1}{j}}\qquad  \textrm{  where $j$ is the number of rows with $\lambda_i^{(s)}\ge i+1$}~.
\end{split}\ee
(See also \eqref{C-anti} for the illustration by Young diagrams.)  
Note that these multiplicity coefficients are still conjectural.

As a simple check, we can compare the result with \cite[(2.29)]{Kim:2018gjo}, which are obtained from SU(4) instanton partition function by ingenious representation theoretic methods. We check the match upto 5-instanton.

\subsection*{New formula of \texorpdfstring{$G_2$}{G2} instanton}

Higssing of $\SO(7)+1\textbf{S}$ leads to pure $G_2$ gauge theory \cite{Kim:2018gjo,Hayashi:2018bkd,Nawata:2021dlk}. Therefore, if we set $m=0$, $A_3=A_1A_2$ in the instanton partition function of $\SO(7)+1\textbf{S}$, we obtain a new formula of $G_2$ instanton partition function. We check it against \cite{Kim:2018gjo,Hayashi:2018bkd,Nawata:2021dlk} upto 4-instanton.

\subsection{\texorpdfstring{$\SO(2N)$}{SO(2N)} with one (conjugate) spinor}

The integral expressions of the partition functions are written in Appendix \ref{app:integral2}.
For a gauge group of $D$ type, the two sectors from $\OO(j)_\pm$ make non-trivial contributions. This is another way to see that the spinor representation and its conjugate are not isomorphic for $\SO(n)$ for general even $n$.

 Hence, the non-trivial JK poles in the unrefined limit are classified $(N+4)$-tuples $\vec\lambda$ of 2d Young diagrams where the numbers of boxes for $Z^{\SO(2N)}_{k,2\ell,+}$  and $Z^{\SO(2N)}_{k,2\ell+1,\pm}$  are subject to
\begin{equation}
    \sum_{s=1}^N |\lambda^{(s)}| = k~,\qquad \sum_{s=N+1}^{N+4} |\lambda^{(s)}| = \ell~,
\end{equation}
whereas those for $Z^{\SO(2N)}_{k,2\ell,-}$  is required to be 
\begin{equation}
    \sum_{s=1}^N |\lambda^{(s)}| = k~,\qquad \sum_{s=N+1}^{N+4} |\lambda^{(s)}| = \ell-1~.
\end{equation}
Then, using the multiplicity coefficients in \eqref{weight2}, the JK residue sum of each sector is expressed as
\begin{footnotesize}
  \begin{equation}
  \begin{split}
        &Z^{\SO(2N)}_{k,2\ell,+}(A_{N+1}=q^{\frac12},A_{N+2}=-q^{\frac12},A_{N+3}=1,A_{N+4}=-1;q)=2^{\text{sign}(\ell)}
        \\
        &\sum_{\vec{\lambda}}C^{\Sp}_{\vec{\lambda},\vec{A}}\prod_{s=1}^{N}\prod_{x\in \lambda^{(s)}}
        \frac{\sh^{4}(2\phi_{s}(x))}
        {\prod\limits_{t=1}^{N} \sh^2(N_{s,t}(x))\, \sh^2(\phi_{s}(x)+ a_t)} \prod_{s\leq t}^{N}\prod_{\substack{x\in \lambda^{(s)}\\y\in \lambda^{(t)}\\ x< y }}\frac{\sh^4(\phi_{s}(x)+\phi_{t}(y))}
        { \sh^2(\phi_{s}(x)+\phi_{t}(y)\pm\hbar)}
        \\
        & \prod_{s=N+1}^{N+4}\prod_{x\in \lambda^{(s)}}
        \frac{\sh^{4}(2\phi_{s}(x))\prod\limits_{t=1}^{N}\sh(\pm\phi_{s}(x)+a_t)}
        {\prod\limits_{t=N+1}^{N+4} \sh^2(N_{s,t}(x))\, \sh^2(\phi_{s}(x)+ a_t)} \prod_{N+1\leq s\leq t}^{N+4}\prod_{\substack{x\in \lambda^{(s)}\\y\in \lambda^{(t)}\\ x< y }}\frac{\sh^4(\phi_{s}(x)+\phi_{t}(y))}
        { \sh^2(\phi_{s}(x)+\phi_{t}(y)\pm\hbar)}
        \\
        & \prod_{s=1}^{N}\prod_{x\in \lambda^{(s)}} \prod_{t=N+1}^{N+4}\prod_{y\in \lambda^{(t)}}\frac{\sh(\hbar\pm\phi_{s}(x)\pm\phi_{t}(y))}
        {\sh(\pm\phi_{s}(x)\pm\phi_{t}(y))}
    \end{split}
\end{equation}
  \begin{equation}
  \begin{split}
        &Z^{\SO(2N)}_{k,2\ell,-}(A_{N+1}=q^{\frac12},A_{N+2}=-q^{\frac12},A_{N+3}=q,A_{N+4}=-q;q)=\Bigg[(-1)^{k}2\frac{\prod_{s=1}^{N}\sh(2a_s)}{\sh^2(\hbar)\sh^2(2\hbar)}\Bigg]^{\text{sign}(\ell)}
        \\
        &\sum_{\vec{\lambda}}C^{\Sp}_{\vec{\lambda},\vec{A}}\prod_{s=1}^{N}\prod_{x\in \lambda^{(s)}}
        \frac{\sh^{4}(2\phi_{s}(x))}
        {\prod\limits_{t=1}^{N} \sh^2(N_{s,t}(x))\, \sh^2(\phi_{s}(x)+ a_t)} \prod_{s\leq t}^{N}\prod_{\substack{x\in \lambda^{(s)}\\y\in \lambda^{(t)}\\x< y }}
        \frac{\sh^4(\phi_{s}(x)+\phi_{t}(y))}
        { \sh^2(\phi_{s}(x)+\phi_{t}(y)\pm\hbar)}
        \\
        & \Bigg[\prod_{s=N+1}^{N+4}\prod_{x\in \lambda^{(s)}}
        \frac{\sh^{4}(2\phi_{s}(x))\prod\limits_{t=1}^{N}\sh(\pm\phi_{s}(x)+a_t)}
        {\prod\limits_{t=N+1}^{N+4} \sh^2(N_{s,t}(x))\, \sh^2(\phi_{s}(x)+ a_t)} \prod_{N+1\leq s\leq t}^{N+4}\prod_{\substack{x\in \lambda^{(s)}\\y\in \lambda^{(t)}\\ x< y }}\frac{\sh^4(\phi_{s}(x)+\phi_{t}(y))}
        { \sh^2(\phi_{s}(x)+\phi_{t}(y)\pm\hbar)}
        \\
         &\prod_{s=1}^{N}\prod_{x\in \lambda^{(s)}}\frac{\sh(2\hbar\pm2\phi_{s}(x))}{\sh^2(2\phi_{s}(x))} \prod_{t=N+1}^{N+4}\prod_{y\in \lambda^{(t)}}\frac{\sh(\hbar\pm\phi_{s}(x)\pm\phi_{t}(y))}
        {\sh(\pm\phi_{s}(x)\pm\phi_{t}(y))}\Bigg]^{\text{sign}(\ell)}
    \end{split}
\end{equation}
  \begin{equation}
  \begin{split}
        &Z^{\SO(2N)}_{k,2\ell+1,+}(A_{N+1}=q^{\frac12},A_{N+2}=-q^{\frac12},A_{N+3}=q,A_{N+4}=-1;q)=(-1)^{k+1}\frac{\prod_{s=1}^{N}\sh(a_s)}{\sh^2(\hbar)}
        \\
        &\sum_{\vec{\lambda}}C^{\Sp}_{\vec{\lambda},\vec{A}}\prod_{s=1}^{N}\prod_{x\in \lambda^{(s)}}
        \frac{\sh^{4}(2\phi_{s}(x))}{\prod\limits_{t=1}^{N} \sh^2(N_{s,t}(x))\, \sh^2(\phi_{s}(x)+ a_t)} \prod_{s\leq t}^{N}\prod_{\substack{x\in \lambda^{(s)}\\y\in \lambda^{(t)}\\ x< y }}\frac{\sh^4(\phi_{s}(x)+\phi_{t}(y))}
        { \sh^2(\phi_{s}(x)+\phi_{t}(y)\pm\hbar)}
        \\
        & \prod_{s=N+1}^{N+4}\prod_{x\in \lambda^{(s)}}
        \frac{\sh^{4}(2\phi_{s}(x))\prod\limits_{t=1}^{N}\sh(\pm\phi_{s}(x)+a_t)}{\prod\limits_{t=N+1}^{N+4} \sh^2(N_{s,t}(x))\, \sh^2(\phi_{s}(x)+ a_t)} \prod_{N+1\leq s\leq t}^{N+4}\prod_{\substack{x\in \lambda^{(s)}\\y\in \lambda^{(t)}\\ x< y }}
        \frac{\sh^4(\phi_{s}(x)+\phi_{t}(y))}
        { \sh^2(\phi_{s}(x)+\phi_{t}(y)\pm\hbar)}
        \\
        &\prod_{s=1}^{N}\prod_{x\in \lambda^{(s)}} \frac{\sh(\hbar\pm\phi_{s}(x))}{\sh^2(\phi_{s}(x))}\prod_{t=N+1}^{N+4}\prod_{y\in \lambda^{(t)}}\frac{\sh(\hbar\pm\phi_{s}(x)\pm\phi_{t}(y))}{\sh(\pm\phi_{s}(x)\pm\phi_{t}(y))}
    \end{split}
\end{equation}
  \begin{equation}
  \begin{split}
        &Z^{\SO(2N)}_{k,2\ell+1,-}(A_{N+1}=q^{\frac12},A_{N+2}=-q^{\frac12},A_{N+3}=1,A_{N+4}=-q;q)=-\frac{\prod_{s=1}^{N}\ch(a_s)}{\sh^2(\hbar)}\\
        &\sum_{\vec{\lambda}}C^{\Sp}_{\vec{\lambda},\vec{A}}\prod_{s=1}^{N}\prod_{x\in \lambda^{(s)}}
        \frac{\sh^{4}(2\phi_{s}(x))}{\prod\limits_{t=1}^{N} \sh^2(N_{s,t}(x))\, \sh^2(\phi_{s}(x)+ a_t)} \prod_{s\leq t}^{N}\prod_{\substack{x\in \lambda^{(s)}\\y\in \lambda^{(t)}\\ x< y }}\frac{\sh^4(\phi_{s}(x)+\phi_{t}(y))}
        { \sh^2(\phi_{s}(x)+\phi_{t}(y)\pm\hbar)}
        \\
        & \prod_{s=N+1}^{N+4}\prod_{x\in \lambda^{(s)}}
        \frac{\sh^{4}(2\phi_{s}(x))\prod\limits_{t=1}^{N}\sh(\pm\phi_{s}(x)+a_t)}{\prod\limits_{t=N+1}^{N+4} \sh^2(N_{s,t}(x))\, \sh^2(\phi_{s}(x)+ a_t)} \prod_{N+1\leq s\leq t}^{N+4}\prod_{\substack{x\in \lambda^{(s)}\\y\in \lambda^{(t)}\\ x< y }}
        \frac{\sh^4(\phi_{s}(x)+\phi_{t}(y))}
        { \sh^2(\phi_{s}(x)+\phi_{t}(y)\pm\hbar)}
        \\
        &\prod_{s=1}^{N}\prod_{x\in \lambda^{(s)}} \frac{\ch(\hbar\pm\phi_{s}(x))}{\ch^2(\phi_{s}(x))}\prod_{t=N+1}^{N+4}\prod_{y\in \lambda^{(t)}}\frac{\sh(\hbar\pm\phi_{s}(x)\pm\phi_{t}(y))}{\sh(\pm\phi_{s}(x)\pm\phi_{t}(y))}
      \end{split}
    \end{equation}
\end{footnotesize}

\subsection{Isomorphisms of representations}

In the cases of $\SO(4)$, $\SO(5)$ and $\SO(6)$, the (conjugate) spinor representation is isomorphic to the fundamental representation of $\SU(2)\times \SU(2)$, $\Sp(2)$ and $\SU(4)$, respectively. This can be seen at the level of the unrefined instanton partition functions as follows, and we check the following identities up to 5-instanton.

\bigskip

\noindent$\bullet\quad(\frakso(4),\mathbf{S})\cong (\fraksu(2)\oplus\fraksu(2),\mathbf{F}\oplus\emptyset)$
\be \widehat Z_{\SO(4)}^{\textrm{spinor}}=Z_{\SU(2),\kappa=\frac12}^{\textrm{fund}} (A \to \sqrt{{A_{1}}{A_{2}}},\mathfrak q\to -\mathfrak q M^{-1/2})Z_{\SU(2),\kappa=0}^{\textrm{pure}}(A \to \sqrt{\frac{A_{1}}{A_{2}}}) Z_{\U(1)}~,\ee
$\bullet\quad(\frakso(4),\mathbf{C})\cong (\fraksu(2)\oplus\fraksu(2),\emptyset\oplus\mathbf{F})$
\be\widehat Z_{\SO(4)}^{\textrm{conj}}=Z_{\SU(2),\kappa=0}^{\textrm{pure}} (A \to \sqrt{{A_{1}}{A_{2}}})Z_{\SU(2),\kappa=\frac12}^{\textrm{fund}}(A \to \sqrt{\frac{A_{1}}{A_{2}}},\mathfrak q\to -\mathfrak q M^{-1/2}) Z_{\U(1)}~,\ee
$\bullet\quad(\frakso(5),\mathbf{S})\cong (\fraksp(2),\mathbf{F})$  \be \label{SO5=Sp2}M^{\frac{k}{2}}\widehat Z_{\SO(5),k}^{\textrm{spinor}}=Z_{\Sp(2),k}^{\textrm{fund}} (A_{1} \to \sqrt{A_{1} A_{2}}, A_{2} \to \sqrt{\frac{A_{1}}{A_{2}}})~,\ee
$\bullet\quad(\frakso(6),\mathbf{S})\cong (\fraksu(4),\mathbf{F})$
\be \label{SO6=SU4} M^{\frac{k}{2}}\widehat Z_{\SO(6),k}^{\textrm{spinor}}=(-1)^k Z_{\SU(4),k,\kappa=\frac12}^{\textrm{fund}}~.\ee
$\bullet\quad(\frakso(6),\mathbf{C})\cong (\fraksu(4),\mathbf{F})$
\be \label{SO6=SU4-2} M^{\frac{k}{2}}\widehat Z_{\SO(6),k}^{\textrm{conj}}=Z_{\SU(4),k,\kappa=-\frac12}^{\textrm{fund}}(M\to M^{-1})~.\ee
Note that the U(1) factor is given by 
\begin{equation}
    \label{U1-instanton} 
    Z_{\U(1)}=\PE\left[- \frac{\frakq\ch \epsilon_{+}}{2 \sh \epsilon_{1,2}}\right]~.
\end{equation}
Also, we need to shift the 5d Chern-Simons level by $\pm\frac{1}{2}$ for $\SU(N)$ gauge theory with one fundamental because of the parity anomaly. 

In particular, \eqref{SO5=Sp2} implies that the top ($j=k$) and bottom ($j=0$) components of the hypermultiplet particles are equal to 
\begin{equation}
\begin{aligned}
    \widehat Z_{k,0}^{\SO(5)}=&Z_{\Sp(2),k,\theta=0 }^{\textrm{pure}}(A_{1} \to \sqrt{A_{1} A_{2}}, A_{2} \to \sqrt{\frac{A_{1}}{A_{2}}}) ~,\cr \widehat Z_{k,k}^{\SO(5)}=&Z_{\Sp(2),k,\theta=\pi }^{\textrm{pure}}(A_{1} \to \sqrt{A_{1} A_{2}}, A_{2} \to \sqrt{\frac{A_{1}}{A_{2}}})~.
\end{aligned}
\end{equation}
Also, \eqref{SO6=SU4} and \eqref{SO6=SU4-2} indicate that
\be
        \widehat Z_{k,0}^{\SO(6)}= (-1)^k Z_{\SU(4),k,\kappa=0 }^{\textrm{pure}} ~,\qquad \widehat Z_{k,k}^{\SO(6)}= Z_{\SU(4),k,\kappa=1 }^{\textrm{pure}}~.\ee

Certainly, of most importance is the triality of $\SO(8)$. The $\SO(8)$ instanton partition functions enjoy the triality \cite{Kim:2019uqw}
\[
\begin{tikzcd}[column sep=1.5em]
 &  (\frakso(8),\textbf{S})  \arrow[dr,"\textrm{c.o.v.}"] \\
 (\frakso(8),\textbf{V} ) \arrow[ur,"\textrm{c.o.v.}"] &&  \arrow[ll,"\textrm{c.o.v.}"] (\frakso(8),\textbf{C})
\end{tikzcd}
\]
where the change of variables for the instanton partition functions is given by
\begin{equation}
A_{1} \to \sqrt{A_{1} A_{2} A_{3} A_{4}}~,\quad A_{2} \to \sqrt{\frac{A_{1} A_{2}}{A_{3} A_{4}}}~,\quad A_{3}\to  \sqrt{\frac{A_{1} A_{3}}{A_{2} A_{4}}}~,\quad A_{4} \to \sqrt{\frac{A_{2} A_{3}}{A_{1} A_{4}}}
\end{equation}
We check the triality upto 5-instanton.

\section{\texorpdfstring{$\SO(n)$ $\cN=1^*$}{SO(N) N=1*} theory}\label{sec:SO-max}
The maximally supersymmetric 5d $\SO(n)$ gauge theory can be realized as a worldvolume theory on D4-branes with an O4${}^-$-plane in Type IIA theory. The scalar field in the adjoint (antisymmetric) hypermultiplet represents the fluctuation of a D4-brane along the transverse direction to its worldvolume. When we turn on the $\Omega$-background transverse to the D4-branes/O4-plane, the adjoint hypermultiplet acquires a mass, giving rise to the $\SO(n)$ $\cN=1^*$ theory.

The equivariant index for the adjoint hypermultiplet is described in detail in \cite{Shadchin:2005mx}. 
The integrand of the JK residue integral for the instanton partition function of the $\SO(n)$ $\cN=1^*$ theory is given by
\begin{footnotesize}
    \begin{align}\label{SO-contour}
  &\cI_{\SO(n),k}^{\textrm{adj}}\cr
  =&\frac{1}{k!2^k} \prod_{I=1}^{k} \frac{\sh{(2\epsilon_+)} \sh(\pm2\phi_I) \sh(\pm2\phi_I+2\e_+)}{ \sh{(\epsilon_{1,2})}  \sh^\chi(\epsilon_{+}\pm \phi_{I}) \, \prod_{s=1}^{\lfloor\frac{n}{2}\rfloor} \sh{ (\pm \phi_{I} \pm a_s + \epsilon_+)} }
  		\prod_{I < J}^{k} \frac{\sh{ ( \pm \phi_I \pm \phi_J)} \sh{ (\pm \phi_{I} \pm \phi_{J} + 2\epsilon_+) }}{\sh{ (\pm \phi_{I} \pm \phi_{J} +\epsilon_{1,2})}}\cr
&\cdot       \prod_{I=1}^{k} \frac{\sh{(\pm m - \epsilon_-)} \sh^\chi(m\pm \phi_{I})  \prod_{s=1}^{\lfloor\frac{n}{2}\rfloor} \sh{(\pm\phi_I \pm a_s + m)} }{\sh{(\pm m - \epsilon_+)} \sh{(\pm2\phi_I\pm m - \epsilon_+)} } \prod_{I < J}^{k} \frac{\sh{(\pm\phi_I \pm \phi_J \pm m - \epsilon_-)}}{\sh{(\pm\phi_I \pm \phi_J \pm m - \epsilon_+)}}
\end{align}
\end{footnotesize}
where $n\equiv\chi  \bmod 2$.

At the unrefined level, $\hbar$ and $m$ play a similar role in the JK residue integral. This is natural because they are the parameters for the $\Omega$-background. Indeed, poles are classified by $(\lfloor \frac{n}{2}\rfloor +4)$-tuples of 4d young diagrams with the total number of boxes $k$.
We specify a pole location associated with a content $x=(x_1,x_2,x_3,x_4)\in \lambda^{(s)}$ as
\begin{equation}\label{4dphi(s)}
  \phi_s(x)= a_s+(x_1-x_2)\hbar+(x_3-x_4)m ~.
\end{equation}
Note that $a_s$ ($s=1,\ldots, \lfloor \frac{n}{2}\rfloor$) are the Coulomb branch parameters of the $\SO(n)$ gauge theory. Similarly to \eqref{additional-Coulomb}, the other four effective Coulomb branch parameters are given by
\be \label{additional-Coulomb-2}
a_{\lfloor \frac{n}{2}\rfloor+j}= \frac{m}{2} (+\pi i )~, \   \frac{\hbar-m}{2} (+\pi i )~,
\ee
for $j=1,2,3,4$.
For the purpose of later use, we define 
\begin{footnotesize}
  \begin{align}\label{4d-YT}
&\wt Z_{k}(\vec{A};q)\cr =&\sum_{\vec{\lambda}}C^{\textrm{4d}}_{\vec{\lambda},\vec{A}}\Bigg(\frac{\sh(m\pm\hbar)\,\sh^2(0)}{\sh^2(\hbar)\,\sh^2(m)} \Bigg)^k
\\&\prod_{s=1}^{\|\vec{\lambda}\|}\prod_{x\in \lambda^{(s)}}\frac{\sh^4(2\phi_s(x))\,\sh^2(2\phi_s(x)+m-\hbar)\, \sh^2(2\phi_s(x)-m+\hbar)\prod_{t=1}^N\sh(\phi_s(x)\pm a_t\pm m)}{\prod_{t=1}^{\|\vec{\lambda}\|}\sh^2(\phi_s(x)\pm a_t)}
\cr&\prod_{s\leq t}^{\|\vec{\lambda}\|}\prod_{\substack{x\in \lambda^{(s)},\,y\in \lambda^{(t)}\\ x< y }}\frac{\sh^4(\phi_s(x)\pm\phi_t(y))\,\sh(\phi_s(x)\pm\phi_t(y)\pm m\pm \hbar)}{\sh^2(\phi_s(x)\pm\phi_t(y)\pm \hbar)\,\sh^2(\phi_s(x)\pm\phi_t(y)\pm m)}\nonumber
\end{align}
\end{footnotesize}
where $\|\vec{\lambda}\|$ is the number of 4d Young diagrams in $\vec\lambda$, and the total ordering on the boxes is the natural extension of \eqref{total-ordering}. In this case, $\|\vec{\lambda}\|=\lfloor \frac{n}{2}\rfloor +4$, and the instanton partition function is
\begin{equation}\label{SO-adj}
    Z_{\SO(n),k}^{\textrm{adj}}=\wt Z_{k}(\vec{A},A_{N+1}=e^{-\frac{m}{2}},A_{N+2}=-e^{-\frac{m}{2}},A_{N+3}=e^{-\frac{\hbar-m}{2}},A_{N+4}=-e^{-\frac{\hbar-m}{2}};q)
\end{equation}
where $N=\lfloor \frac{n}{2}\rfloor$.
As in \eqref{Dweight}, the non-trivial multiplicity constants
are involved due to higher-order poles coming from the additional effective Coulomb branch parameters \eqref{additional-Coulomb-2}. We do \emph{not} know how to determine these multiplicity constants for general 4d Young diagrams. However, some of the multiplicity constants $C^{\textrm{4d}}_{\vec{\lambda},\vec{A}}$ are listed in Appendix \ref{app:coefficient}.

Since the adjoint representation $\boldsymbol{6}$ of $\mathfrak{so}(4)$ is isomorphic to the direct sum $(\boldsymbol{3,1})\oplus (\boldsymbol{1,3})$ of the adjoint representations of $\mathfrak{su}(2)\oplus \mathfrak{su}(2)$, we have the following identity at the unrefined level
\begin{equation}
    Z_{\SO(4)}^{\textrm{adj}}(A_1,A_2,M)=    Z_{\SU(2)}^{\textrm{adj}}(A=(A_1A_2)^{\frac12},M) Z_{\SU(2)}^{\textrm{adj}}(A=(A_1/A_2)^{\frac12},M) Z_{\textrm{extra}}
\end{equation}
where
\begin{equation}
     Z_{\textrm{extra}}=\mathrm{PE}\left[ \frac{\frakq(2+3 M+2 M^{2})(q-M)(1-M q)}{(1-\frakq)M(1+M)^2(1-q)^2}\right]~.
\end{equation}
Also, the adjoint representation of $\mathfrak{so}(6)$ is isomorphic to that of $\mathfrak{su}(4)$, which leads to
\begin{equation}
    Z_{\SO(6)}^{\textrm{adj}}(A_1,A_2,A_3,M)= Z_{\SU(4)}^{\textrm{adj}}(A_1,A_2,A_3,M) Z^\prime_{\textrm{extra}}
\end{equation}
where
\begin{equation}
     Z^\prime_{\textrm{extra}}=\mathrm{PE}\left[ \frac{\frakq(1+ M+M^{2})(q-M)(1-M q)}{(1-\frakq)M(1+M)^2(1-q)^2}\right]~.
\end{equation}
We check these identities upto 5-instanton.

\section{\texorpdfstring{$\Sp(N)$}{Sp(N)} gauge group with antisymmetric hypermultiplet}\label{sec:Sp-anti}

If we replace the O4${}^-$-plane by an O8${}^-$-plane in the brane configuration described in the previous section, the gauge group becomes $\Sp(N)$, and the fluctuation of a D4-brane along the transverse directions to its worldvolume on the O8${}^-$-plane describes the scalar field in the antisymmetric hypermultiplet. Hence, the D4-O8${}^-$ system in Type IIA theory gives rise to the $\Sp(N)$ gauge theory with the antisymmetric hypermultiplet.

This brane system has close connections to Type I' theory. From the viewpoint of Type I' theory, the instanton quantum mechanics is described in \cite[Appendix D]{Kim:2012gu}.
Using the instanton quantum mechanics, we can write down the instanton partition functions as JK contour integral formulas \cite{Shadchin:2005mx,Kim:2012gu,Hwang:2014uwa}. The results are summarized in Appendix \ref{app:integral3}.

Again, the gauge group $\OO(k)$ of the instanton quantum mechanics has two components, which results in two contributions $Z^{\textrm{anti}}_{\Sp(N),k,\pm}$ to the partition function. Then, the $\theta$-angle can be distinguished through the following equations:
\begin{equation}
Z_{\Sp(N),k,\theta=0}^{\textrm{anti}}=\frac{Z^{\textrm{anti}}_{\Sp(N),k,+}+Z^{\textrm{anti}}_{\Sp(N),k,-}}{2}~,\quad  Z_{\Sp(N),k,\theta=\pi}^{\textrm{anti}}=(-1)^k\frac{Z^{\textrm{anti}}_{\Sp(N),k,+}-Z^{\textrm{anti}}_{\Sp(N),k,-}}{2}~.
\end{equation}

In the unrefined limit, the JK poles are classified using 4d Young diagrams, and a pole location at a content $x\in\lambda^{(s)}$ is as in \eqref{4dphi(s)}.
Additionally, all the residues can be expressed in a single universal formula, as shown in (\ref{4d-YT}). This is similar to the relationship between the instanton partition functions of pure Yang-Mills theory for $\SO(n)$ and $\Sp(N)$ gauge groups \cite{Nawata:2021dlk}. This relationship still holds when including an antisymmetric hypermultiplet. However, there are more effective Coulomb branch parameters than  \eqref{SO-adj}, depending on the instanton number and plus/minus sector:

\bigskip

\noindent $\bullet$ (even,$+$) sector
\begin{equation}\label{Sp-even-plus}
    Z^{\textrm{anti}}_{\Sp(N),k=2\ell,+}=\wt Z_\ell(\vec{A};q)
\end{equation}
where there are 8 additional effective Coulomb branch parameters ($j=1,\ldots,8$)
$$    a_{N+j}=  \frac{\hbar}{2} (+\pi i )~, \   \frac{m}{2} (+\pi i )~, \   \frac{\hbar-m}{2} (+\pi i )~, \ 0(+\pi i )~.$$

\noindent $\bullet$ (odd,$+$) sector
\begin{equation}\label{Sp-odd-plus}
Z^{\textrm{anti}}_{\Sp(N),k=2\ell+1,+}=\frac1{2\sh^2(\hbar)\sh^2(m)}\prod^N_{t=1}\frac{\sh(a_t\pm m)}{\sh^2(a_t)} \prod_{s=1}^{\|\vec{\lambda}\|}\prod_{x\in \lambda^{(s)}}\sh(\phi_s(x)\pm m\pm\hbar)\wt Z_\ell(\vec{A};q)
\end{equation}
where there are 9 additional effective Coulomb branch parameters ($j=1,\ldots,9$)
$$    a_{N+j}=  \frac{\hbar}{2} (+\pi i )~, \   \frac{m}{2} (+\pi i )~, \   \frac{\hbar-m}{2} (+\pi i )~, \ \pi i ~, \ \hbar~, \ m~.$$

\noindent $\bullet$ (even,$-$) sector
\begin{equation}\label{Sp-even-minus}
Z^{\textrm{anti}}_{\Sp(N),k=2\ell,+}=\frac{\ch(m\pm\hbar)}{\sh^2(\hbar)\sh^2(2\hbar)\sh^2(m)\sh^2(2m)}\prod^N_{t=1}\frac{\sh(a_t\pm m)\ch(a_t\pm m)}{\sh^2(2a_t)}\wt Z_{\ell-1}(\vec{A};q)
\end{equation}
where there are 10 additional effective Coulomb branch parameters ($j=1,\ldots,10$)
$$    a_{N+j}=\frac{\hbar}{2} (+\pi i )~, \   \frac{m}{2} (+\pi i )~, \   \frac{\hbar-m}{2} (+\pi i )~, \ \hbar (+\pi i )~, \ m (+\pi i )~.$$

\noindent $\bullet$ (odd,$-$) sector
\begin{equation}\label{Sp-odd-minus}
Z^{\textrm{anti}}_{\Sp(N),k=2\ell+1,+}=\frac1{2\sh^2(\hbar)\sh^2(m)}\prod^N_{t=1}\frac{\ch(a_t\pm m)}{\ch^2(a_t)}\prod_{s=1}^{\|\vec{\lambda}\|}\prod_{x\in \lambda^{(s)}}\ch(\phi_s(x)\pm m\pm\hbar) \wt Z_\ell(\vec{A};q)
\end{equation}
where there are 9 additional effective Coulomb branch parameters  ($j=1,\ldots,9$)
$$    a_{N+j}=   \frac{\hbar}{2} (+\pi i )~, \   \frac{m}{2} (+\pi i )~, \   \frac{\hbar-m}{2} (+\pi i )~, \ 0 ~, \ \hbar+\pi i~, \ m+\pi i~.$$

Note that  the multiplicity coefficients are involved for the additional Coulomb branch parameters due to the presence of higher-order poles in \eqref{4d-YT}, and we do not know how to determine them for general 4d Young diagrams. Nonetheless, we list some of them in Appendix \ref{app:coefficient}.

We can verify the isomorphisms of representations by comparing the instanton partition functions of different theories. For example, the antisymmetric representation of $\Sp(1)$ is trivial, so it decouples from the gauge theory and does not affect the instanton dynamics. Thus, the instanton partition function for the $\Sp(1)$ theory with an antisymmetric hypermultiplet is equal to the pure $\Sp(1)$ instanton partition function
\begin{align}
    \label{Sp1Sp1}
  Z_{\Sp(1),\theta=0}^{\textrm{anti}}=&  Z_{\Sp(1),\theta=0}^{\textrm{pure}} Z_{\textrm{D0}}~,\cr Z_{\Sp(1),\theta=\pi}^{\textrm{anti}}=& Z_{\Sp(1),\theta=\pi}^{\textrm{pure}}
\end{align}
up to the 8 translational zero modes of a single D0-brane
\be
Z_{\textrm{D0}}= \PE \Big[\frac{\frakq}{ \sh ({\epsilon_{1,2}}) \sh ({m\pm \e_+})}\Big]~.
\ee
On the other hand, the antisymmetric representation of $\Sp(2)$ is isomorphic to the vector representation of $\SO(5)$, so the instanton partition functions should obey the relation 
\begin{equation}\label{Sp2So5}
  Z_{\Sp(2),\theta=0}^{\textrm{anti}}Z_{\U(1)}=  Z_{\SO(5)}^{\textrm{vect}}Z_{\textrm{D0}}
\end{equation}
where the U(1) instanton partition function is given in \eqref{U1-instanton}.
We have checked \eqref{Sp1Sp1} and \eqref{Sp2So5} upto 8-instanton at the unrefined level, respectively.

\begin{figure}[ht]
    \centering
   \includegraphics[width=5.5cm]{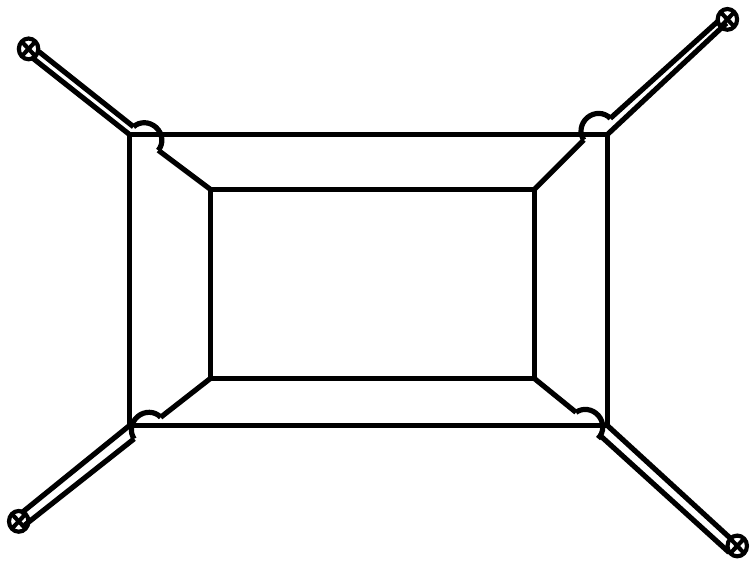}\qquad     \includegraphics[width=7cm]{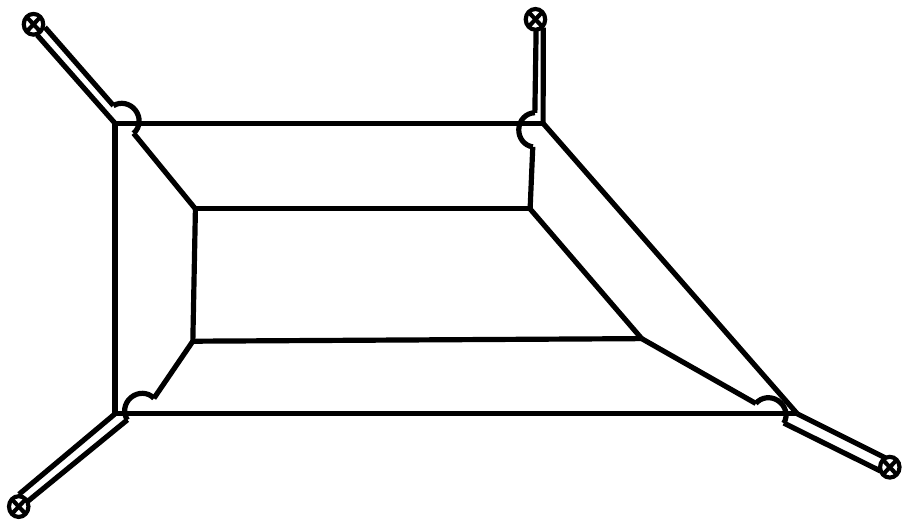}
    \caption{Fivebrane web configuration of $\Sp(2)$  gauge theory with massless antisymmetric hypermultiplet in Type IIB theory where the left-panel represents $\theta=0$ while the right-panel illustrates $\theta=\pi$. }
    \label{fig:Sp2-anti-sym}
\end{figure}

Once we take T-duality from the brane system in Type IIA theory, we can construct
5d $\Sp(N)$ theories with an antisymmetric hypermultiplet by fivebrane web with two O7${}^-$-planes in Type IIB theory  \cite{Bergman:2015dpa,Hayashi:2018lyv,Cheng:2018wll,Hayashi:2019yxj}. The brane configuration can be further deformed by splitting an O7${}^-$-plane into two $(p,q)$-sevenbranes \cite{Sen:1996vd} and performing Hanany-Witten transitions \cite{Hanany:1996ie}. In the massless limit of the antisymmetric matter ($m= \e_+$ at the refined level), the resulting fivebrane web takes the form of a product of pure $\Sp(1)$ theories \cite{Cheng:2018wll}, as shown in Figure \ref{fig:Sp2-anti-sym}. This brane moves and the Higgsing procedure can also be verified at the level of instanton partition functions as
\begin{align}
  Z_{\Sp(2),\theta=0}^{\textrm{anti}}(m=\e_+)=&  Z_{\Sp(1),\theta=0}^{\textrm{pure}}(A_1)Z_{\Sp(1),\theta=0}^{\textrm{pure}}(A_2)Z_{\textrm{D0}}\cr
    Z_{\Sp(2),\theta=\pi}^{\textrm{anti}}(m=\e_+)=& Z_{\Sp(1),\theta=\pi}^{\textrm{pure}}(A_1)Z_{\Sp(1),\theta=\pi}^{\textrm{pure}}(A_2)~.
\end{align}

In order to move forward with the research presented in this paper, it is important to fully determine the multiplicity constants in \eqref{4d-YT}. One potential approach to this problem is to consider the blowup equation on $\bC^2_{\pm\hbar}\times \bC^2_{\pm m}$, as the equivariant parameters $\hbar$ and $m$ play a similar role in this context.

While \eqref{4d-YT} currently yields non-trivial rational functions through the cancellation of Sh(0) in the numerator with poles in the denominator, it would be valuable to find a closed-form expression similar to the Nekrasov factor \eqref{Nij} for 4d Young diagrams. This could provide deeper insights and simplify calculations furthermore.

\acknowledgments

We are grateful to Sung-Soo Kim, Xiaobin Li, and Futoshi Yagi for the valuable discussion on the related topic \cite{Chengdu}.  Their help was essential in the development of this paper, particularly the results in \S\ref{sec:SU-sym} which stemmed from the collaboration with them. 
We would also like to thank Jiaqi Guo and Rui-Dong Zhu for the discussion. Additionally, we extend our sincere gratitude to the anonymous referee of JHEP who carefully reviewed our paper. S.N. is indebted to Southwest Jiaotong University for the hospitality where a part of the work is carried out. The research of S.N. is supported by the National Science Foundation of China under Grant No.12050410234 and Shanghai Foreign Expert grant No. 22WZ2502100.

\appendix

\section{Notations and definitions}\label{app:notations}
In this appendix, we provide a summary of the notations and definitions used throughout the paper.

We denote a 2d Young diagram by $\lambda=\left(\lambda_1,\lambda_2,\cdots\right)$, which is a sequence of non-negative integers such that $\lambda_i \geq \lambda_{i+1}$ and $|\lambda|=\sum_i\lambda_i<\infty$. We write the transposition of $\lambda$ by $\lambda^t$.  The arm length $a_{\lambda}(x)$ of a box at $x=(x_1,x_2)$ in the Young diagram is the number of boxes to the right of the box, and the leg length $l_{\lambda}(x)$ is the number of boxes below it. This is illustrated in the following figure:
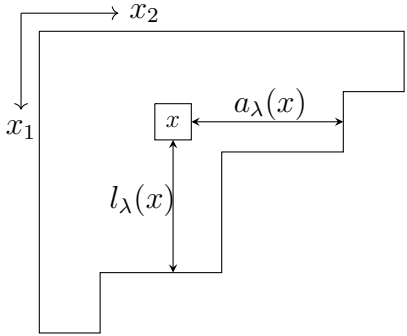
\begin{figure}[ht]\centering
  \begin{tikzpicture}[scale=0.8]
  \draw[->]   (-.3,.3)--(1.3,.3) node[right] {$x_2$};
  \draw[->]   (-.3,.3)--(-.3,-1.3) node[below] {$x_1$};
\draw (0,0)--(0,-5)--(1,-5)--(1,-4)--(3,-4)--(3,-2)--(5,-2)--(5,-1)--(6,-1)--(6,0)--(0,0);
\draw (1.9,-1.2)--(1.9,-1.8)--(2.5,-1.8)--(2.5,-1.2)--(1.9,-1.2);
\draw [<->,>=stealth] (2.5,-1.5)--(5,-1.5);
\node [scale=0.8] at (2.2,-1.5)  {$x$};
\draw [<->,>=stealth] (2.2,-1.8)--(2.2,-4);
\draw (3.8,-1.2) node [scale=1] {$a_{\lambda}(x)$};
\draw (1.7,-2.8) node [scale=1] {$l_{\lambda}(x)$};
  \end{tikzpicture}
  \caption{Arm and leg length of a box $x=(x_1,x_2)$ in a Young diagram.}
\end{figure}

We also introduce the notation for 5d instanton partition functions. 
The moduli spaces of instantons  receive equivariant actions of $\SO(2)_{\e_1}\times\SO(2)_{\e_1}$ of the space-time. Also, the maximal torus $\prod_{s=1}^{{\textrm{rank}G}}\U(1)_{a_s}$ of the gauge group $G$ and that $\U(1)_{m}$ of the flavor group act on the instanton moduli spaces. Consequently, 5d instanton partition functions are rational functions of variables
$$
q=e^{-\e_1}~, \quad t=e^{\e_2}~,  \quad A_s=e^{- a_s}~, \quad M=e^{- m}.
$$
The parameters $A_s$ are Coulomb branch parameters, and the parameter $M$ is the mass parameter of a hypermultiplet.
We also use the notation $2\e_\pm=\e_1\pm\e_2$.

In this paper, we mainly focus on the unrefined limit $\e_1=-\e_2=\hbar$, and we write  $q=e^{-\hbar}$.
We denote a $k$-instanton partition function with a hypermultiplet in a representation of $G$ by $Z_{G,k}^{\textrm{rep}}$. This paper obtains expressions of $k$-instanton partition functions summed over $P$-tuples of Young diagrams, which is denoted by
$$
\vec{\lambda}=(\lambda^{(1)},\ldots,\lambda^{(P)})~,
$$
where the total number of boxes satisfy
$$
k=\sum_{s=1}^{\|\vec{\lambda}\|}|\lambda^{(s)}|~.
$$
Here we also denote the number of Young diagrams by $\|\vec{\lambda}\|=P$.
The total instanton partition function is expressed as a generating function with instanton counting parameter $\frakq$
$$
Z_G^{\textrm{rep}}=\sum_{k=0}^\infty \frakq^k Z_{G,k}^{\textrm{rep}}(A_i;q)~.
$$
The perturbative partition function is often written in terms of the plethystic exponent
  \be\label{plethystic}
\textrm{PE}\left[f(x,y,\cdots)\right]\equiv\exp\left[\sum_{d=1}^\infty\frac{1}{d}
  f(x^d,y^d,\cdots)\right] \ee  
which brings the single particle index $f$ to the multi-particle index.

With these notations and definitions, we can now proceed with the analysis presented in the main text of the paper.

\section{Integral expressions of instanton partition functions}\label{app:integral}

To be self-contained, in this Appendix,  we provide a summary of the partition functions of supersymmetric quantum mechanics on instanton moduli spaces, as studied in \cite{Kim:2018gjo,Kim:2012gu}. For more details, we refer the reader to the original references.

\subsection{\texorpdfstring{$\SO(2N+1)$ with spinor hypermultiplet}{SO(2N+1) with spinor hypermultiplet}}\label{app:integral1}

\begin{figure}[ht]
    \centering
   \includegraphics[width=5cm]{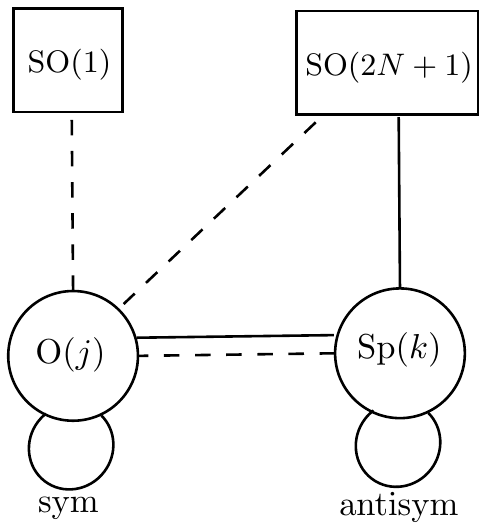}
    \caption{The solid line represents a hypermultiplet while the dashed line represents a Fermi multiplet. }
    \label{fig:ADHM-odd}
\end{figure}

The brane system for $\SO(2N+1)$ gauge theory with spinor hypermultiplet is schematically drawn in Figure \ref{fig:SO-spinor}. The instanton quantum mechanics can be derived from open strings in this brane system. $\cN=4$ supersymmetric quantum mechanics on $k$ D1-branes and $j$ D1'-branes is given in Figure \ref{fig:ADHM-odd}. The formal $\SO(1)$  comes from the half D5-brane on the $\widetilde{\mathrm{O}5}^{+}$-plane, on the leg part of fivebranes in Figure \ref{fig:SO-spinor}. For 5d gauge groups of $B$ type, the contributions from $\OO(2\ell)_-$ and $\OO(2\ell+1)_{+}$ sectors vanish due to the fermionic zero modes. 

Hence, the integrand of the supersymmetric quantum mechanics for $k$ instantons and $2\ell$ hypermultiplet particles is given by
\begin{footnotesize}
  \begin{align}\label{B-even}
        &\cI^{\SO(2N+1)}_{k,2\ell}\cr=&
        \frac{1}{2^{k}\,k!}\frac{\sh^{k}(2\epsilon_{+})}{\sh^{k}(\epsilon_{1,2})}
        \prod_{I=1}^{k}\frac{\sh(2\epsilon_{+}\pm2\phi_{I})\,
        \sh(\pm2\phi_{I})}{\sh(\epsilon_{+}\pm\phi_{I})\prod_{s=1}^{N}\sh(\epsilon_{+}\pm\phi_{I}\pm
        a_s)}  \prod_{I>J}\frac{\sh(2\epsilon_{+}\pm\phi_{I}\pm\phi_{J})\,
        \sh(\pm\phi_{I}\pm\phi_{J})}{\sh(\epsilon_{1,2}\pm\phi_{I}\pm\phi_{J})}
        \cr
        &\frac{1}{2^{\ell}\,\ell!}\frac{\sh^{\ell}(2\epsilon_{+})}{
\sh^\ell(\epsilon_{1,2})}\prod_{i=1}^{\ell}\frac{\sh(\pm\chi_{i})\prod_{s=1}^{N}\sh(\pm\chi_{i}+a_s)}{
        \sh(\epsilon_{1,2}\pm 2\chi_{i})}
        \prod_{i>j}\frac{\sh(2\epsilon_{+}\pm\chi_{i}\pm\chi_{j})
        \,\sh(\pm\chi_{i}\pm\chi_{j})}{
        \sh(\epsilon_{1,2}\pm\chi_{i}\pm\chi_{j})}\cr
        &\prod_{I=1}^{k}
        \prod_{i=1}^{\ell}\frac{\sh(\epsilon_{-}\pm\phi_{I}\pm\chi_{i})}
        {\sh(-\epsilon_{+}\pm
        \phi_{I}\pm\chi_{i})}\ .
    \end{align}
\end{footnotesize}
 Here $\phi_I$ $(I=1,\cdots, k)$ are $\Sp(k)$ gauge fugacities, and $\chi_i$ $(i=1,\cdots,\ell)$ are $\OO(2\ell)$ gauge fugacities so that they are integrated by the JK prescription. 
On the other hand,  the integrand  for $k$ instantons and $(2\ell+1)$ hypermultiplet particles is given by
\begin{footnotesize}
  \begin{align}\label{B-odd}
        &\cI^{\SO(2N+1)}_{k,2\ell+1}\cr=&
        \frac{1}{2^{k}\,k!}\frac{\sh^{k}(2\epsilon_{+})}{\sh^{k}(\epsilon_{1,2})}
        \prod_{I=1}^{k}\frac{\sh(2\epsilon_{+}\pm2\phi_{I})\,
        \sh(\pm2\phi_{I})}{\sh(\epsilon_{+}\pm\phi_{I})\prod_{s=1}^{N}\sh(\epsilon_{+}\pm\phi_{I}\pm
        a_s)}  \prod_{I>J}\frac{\sh(2\epsilon_{+}\pm\phi_{I}\pm\phi_{J})\,
        \sh(\pm\phi_{I}\pm\phi_{J})}{\sh(\epsilon_{1,2}\pm\phi_{I}\pm\phi_{J})}
        \cr
        &\frac{1}{2^{\ell}\,\ell!}\frac{\sh^\ell(2\epsilon_{+})}{\sh^{\ell+1}(\epsilon_{1,2})}\prod_{i=1}^{\ell}\frac{\ch(2\epsilon_{+}\pm\chi_{i})\,
        \sh(\pm2\chi_{i})\prod_{s=1}^{N}\sh(\pm\chi_{i}+a_s)}{\ch(\epsilon_{1,2}\pm\chi_{i})\,
        \sh(\epsilon_{1,2}\pm 2\chi_{i})}\cdot\prod_{s=1}^{N}\ch(a_s)
        \cr
        &\prod_{i>j}\frac{\sh(2\epsilon_{+}\pm\chi_{i}\pm\chi_{j})
        \,\sh(\pm\chi_{i}\pm\chi_{j})}{
        \sh(\epsilon_{1,2}\pm\chi_{i}\pm\chi_{j})}
        \prod_{I=1}^{k}\frac{\ch(\epsilon_{-}\pm\phi_{I})}{\ch(-\epsilon_{+}\pm\phi_{I})}\prod_{i=1}^{\ell}
        \frac{\sh(\epsilon_{-}\pm\phi_{I}\pm\chi_{i})}{
        \sh(-\epsilon_{+}\pm\phi_{I}\pm\chi_{i})}\ .
    \end{align}
\end{footnotesize}

\subsection{\texorpdfstring{$\SO(2N)$ with (conjugate) spinor hypermultiplet}{SO(2N)  with (conjugate) spinor hypermultiplet}}\label{app:integral2}

\begin{figure}[ht]
    \centering
   \includegraphics[width=5cm]{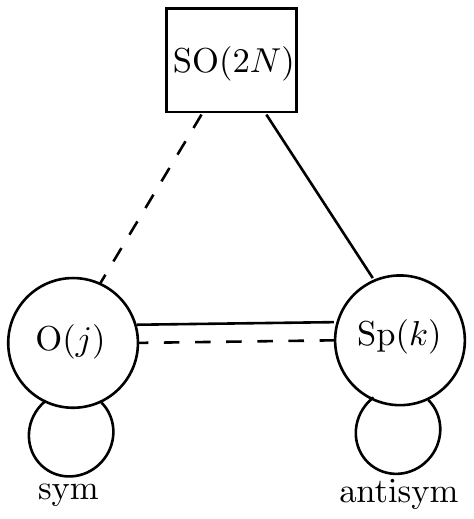}
    \caption{The solid line represents a hypermultiplet while the dashed line represents a Fermi multiplet. }
    \label{fig:ADHM-even}
\end{figure}

The brane system for $\SO(2N)$ gauge theory with (conjugate) spinor hypermultiplet is shown in Figure \ref{fig:SO-spinor}. This brane system allows us to derive the instanton quantum mechanics from open strings. In Figure \ref{fig:ADHM-even}, we see the $\cN=4$ supersymmetric quantum mechanics on $k$ D1-branes and $j$ D1'-branes. In this case, there are non-trivial contributions from the two connected components of the gauge group $\OO(j)$.

To account for each contribution, we write down an integrand of the partition function for each one:
\begin{footnotesize}
  \begin{equation}\label{D-even-p}
  \begin{split}
        &\cI^{\SO(2N)}_{k,2\ell,+}=
        \\
        &\frac{1}{2^{k}\,k!}\frac{\sh^{k}(2\epsilon_{+})}{\sh^{k}(\epsilon_{1,2})}\prod_{I=1}^{k}\frac{\sh(2\epsilon_{+}\pm2\phi_{I})\,\sh(\pm2\phi_{I})}{\prod\limits_{s=1}^{N}\sh(\epsilon_{+}\pm\phi_{I}\pm a_s)}\prod_{I>J}\frac{\sh(2\epsilon_{+}\pm\phi_{I}\pm\phi_{J})\,\sh(\pm\phi_{I}\pm\phi_{J})}{\sh(\epsilon_{1,2}\pm\phi_{I}\pm\phi_{J})}
        \\
        &\frac{2^{\text{sign}(\ell)}}{2^{\ell}\,\ell!}\frac{\sh^\ell(2\epsilon_{+})}{\sh^\ell(\epsilon_{1,2})}\prod_{i=1}^{\ell}\frac{\prod\limits_{s=1}^{N}\sh(\pm\chi_{i}+a_s)}{
        \sh(\epsilon_{1,2}\pm 2\chi_{i})}
        \prod_{i>j}\frac{\sh(2\epsilon_{+}\pm\chi_{i}\pm\chi_{j})\,\sh(\pm\chi_{i}\pm\chi_{j})}{
        \sh(\epsilon_{1,2}\pm\chi_{i}\pm\chi_{j})}
        \\
        &\prod_{I=1}^{k}\prod_{i=1}^{\ell}\frac{\sh(\epsilon_{-}\pm\phi_{I}\pm\chi_{i})}{\sh(-\epsilon_{+}\pm
        \phi_{I}\pm\chi_{i})}
    \end{split}
\end{equation}
\begin{equation}\label{D-even-m}
  \begin{split}
        &\cI^{\SO(2N)}_{k,2\ell,-}=
        \\
        &\frac{1}{2^{k}\,k!}\frac{\sh^{k}(2\epsilon_{+})}{\sh^{k}(\epsilon_{1,2})}  \prod_{I=1}^{k}\frac{\sh(2\epsilon_{+}\pm2\phi_{I})\,
        \sh(\pm2\phi_{I})}{\prod\limits_{s=1}^{N}\sh(\epsilon_{+}\pm\phi_{I}\pm a_s)}\prod_{I>J}\frac{\sh(2\epsilon_{+}\pm\phi_{I}\pm\phi_{J})\,\sh(\pm\phi_{I}\pm\phi_{J})}{\sh(\epsilon_{1,2}\pm\phi_{I}\pm\phi_{J})}
        \\
        &\Bigg[\frac{\prod\limits_{s=1}^{N}\sh(2a_s)}{2^{\ell-1}\,(\ell-1)!}\frac{\ch(2\epsilon_{+})}{\sh(2\epsilon_{1,2})\sh(\epsilon_{1,2})}\prod_{i=1}^{\ell-1}\frac{\sh(2\epsilon_{+})\,\sh(4\epsilon_{+}\pm2\chi_{i})\,\sh(\pm2\chi_{i})\prod\limits_{s=1}^{N}\sh(\pm\chi_{i}+a_s)}{\sh(\epsilon_{1,2})\,\sh(2\epsilon_{1,2}\pm2\chi_{i})\,\sh(\epsilon_{1,2}\pm 2\chi_{i})}
        \\
        &\prod_{i>j}\frac{\sh(2\epsilon_{+}\pm\chi_{i}\pm\chi_{j})\,\sh(\pm\chi_{i}\pm\chi_{j})}{
        \sh(\epsilon_{1,2}\pm\chi_{i}\pm\chi_{j})}\prod_{I=1}^{k}\frac{\sh(2\epsilon_{-}\pm2\phi_{I})}{\sh(-2\epsilon_{+}\pm2\phi_{I})}\prod_{i=1}^{\ell-1}\frac{\sh(\epsilon_{-}\pm\phi_{I}\pm\chi_{i})}{\sh(-\epsilon_{+}\pm\phi_{I}\pm\chi_{i})}\Bigg]^{\text{sign}(\ell)}
    \end{split}
\end{equation}
\begin{equation}\label{D-odd-p}
  \begin{split}
        &\cI^{\SO(2N)}_{k,2\ell+1,+}=
        \\
        &\frac{1}{2^{k}\,k!}\frac{\sh^{k}(2\epsilon_{+})}{\sh^{k}(\epsilon_{1,2})}
        \prod_{I=1}^{k}\frac{\sh(2\epsilon_{+}\pm2\phi_{I})\,
        \sh(\pm2\phi_{I})}{\prod\limits_{s=1}^{N}\sh(\epsilon_{+}\pm\phi_{I}\pm a_s)}  \prod_{I>J}\frac{\sh(2\epsilon_{+}\pm\phi_{I}\pm\phi_{J})\,\sh(\pm\phi_{I}\pm\phi_{J})}{\sh(\epsilon_{1,2}\pm\phi_{I}\pm\phi_{J})}
        \\
        &\frac{\prod\limits_{s=1}^{N}\sh(a_s)}{2^{\ell}\,\ell!}\frac{\sh^\ell(2\epsilon_{+})}{\sh^{\ell+1}(\epsilon_{1,2})}\prod_{i=1}^{\ell}\frac{\sh(2\epsilon_{+}\pm\chi_{i})\,
        \sh(\pm\chi_{i})\prod\limits_{s=1}^{N}\sh(\pm\chi_{i}+a_s)}{\sh(\epsilon_{1,2}\pm\chi_{i})\,
        \sh(\epsilon_{1,2}\pm 2\chi_{i})}
        \\
        &\prod_{i>j}\frac{\sh(2\epsilon_{+}\pm\chi_{i}\pm\chi_{j})\,\sh(\pm\chi_{i}\pm\chi_{j})}{
        \sh(\epsilon_{1,2}\pm\chi_{i}\pm\chi_{j})}
        \prod_{I=1}^{k}\frac{\sh(\epsilon_{-}\pm\phi_{I})}{\sh(-\epsilon_{+}\pm\phi_{I})}\prod_{i=1}^{\ell}
        \frac{\sh(\epsilon_{-}\pm\phi_{I}\pm\chi_{i})}{\sh(-\epsilon_{+}\pm\phi_{I}\pm\chi_{i})}
    \end{split}
\end{equation}
  \begin{equation}\label{D-odd-m}
  \begin{split}
        &\cI^{\SO(2N)}_{k,2\ell+1,-}=
        \\
        &\frac{1}{2^{k}\,k!}\frac{\sh^{k}(2\epsilon_{+})}{\sh^{k}(\epsilon_{1,2})}
        \prod_{I=1}^{k}\frac{\sh(2\epsilon_{+}\pm2\phi_{I})\,
        \sh(\pm2\phi_{I})}{\prod\limits_{s=1}^{N}\sh(\epsilon_{+}\pm\phi_{I}\pm a_s)}\prod_{I>J}\frac{\sh(2\epsilon_{+}\pm\phi_{I}\pm\phi_{J})\,
        \sh(\pm\phi_{I}\pm\phi_{J})}{\sh(\epsilon_{1,2}\pm\phi_{I}\pm\phi_{J})}
        \\
        &\frac{\prod\limits_{s=1}^{N}\ch(a_s)}{2^{\ell}\,\ell!}\frac{\sh^\ell(2\epsilon_{+})}{\sh^{\ell+1}(\epsilon_{1,2})}\prod_{i=1}^{\ell}\frac{\ch(2\epsilon_{+}\pm\chi_{i})\,
        \ch(\pm\chi_{i})\prod\limits_{s=1}^{N}\sh(\pm\chi_{i}+a_s)}{\ch(\epsilon_{1,2}\pm\chi_{i})\,
        \sh(\epsilon_{1,2}\pm 2\chi_{i})}
        \\
        &\prod_{i>j}\frac{\sh(2\epsilon_{+}\pm\chi_{i}\pm\chi_{j})
        \,\sh(\pm\chi_{i}\pm\chi_{j})}{
        \sh(\epsilon_{1,2}\pm\chi_{i}\pm\chi_{j})}
        \prod_{I=1}^{k}\frac{\ch(\epsilon_{-}\pm\phi_{I})}{\ch(-\epsilon_{+}\pm\phi_{I})}\prod_{i=1}^{\ell}
        \frac{\sh(\epsilon_{-}\pm\phi_{I}\pm\chi_{i})}{
        \sh(-\epsilon_{+}\pm\phi_{I}\pm\chi_{i})}
    \end{split}
\end{equation}
\end{footnotesize}

\subsection{\texorpdfstring{$\Sp(N)$}{Sp(N)} with antisymmetric hypermultiplet}\label{app:integral3}

Let us recall the supersymmetric quantum mechanics of the $k$ instanton moduli space of pure $\Sp(N)$ gauge theory \cite{Nekrasov-Shadchin}. This system is described by an $\OO(k)$ gauge theory with a (rank-two) symmetric hypermultiplet and $2N$ fundamental half-hypermultiplets, which have $\Sp(N)$ flavor symmetry.

\begin{figure}[ht]
    \centering
   \includegraphics[width=2cm]{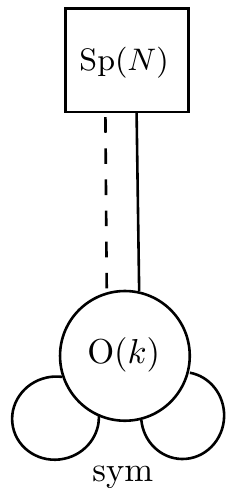}
    \caption{The solid line represents a hypermultiplet while the dashed line represents a Fermi multiplet. }
    \label{fig:ADHM-anti}
\end{figure}

The introduction of a 5d antisymmetric hypermultiplet modifies the field content of the instanton quantum mechanics. Specifically, it introduces an additional symmetric hypermultiplet and $2N$ fundamental half-Fermi multiplets. (This is illustrated in Figure \ref{fig:ADHM-anti}.) The equivariant index for the 5d antisymmetric hypermultiplet has previously been calculated in \cite[(5.14)]{Shadchin:2005mx}. For more details on the field content of the instanton quantum mechanics, see \cite[Appendix D]{Kim:2012gu}.

The partition function receives two contributions due to the $\OO(k)$ gauge group. These contributions can be calculated by performing the JK residue integrals of the following integrands:
\begin{footnotesize}
\begin{align}
  &\cI_{k=2\ell+\chi}^+\cr=&
	    \left(\frac{1}{\sh{(\epsilon_{1,2})} \, \prod_{s=1}^{N} \sh{(\pm a_s + \epsilon_+)}} \cdot \prod_{I=1}^{\ell} \frac{ \sh{(\pm \phi_I)}\sh{ (\pm \phi_I + 2\epsilon_+)}} {\sh{ (\pm \phi_I +\epsilon_{1,2})}}\right)^{\chi}\\
&\cdot \prod_{I=1}^{\ell} \frac{\sh{(2\epsilon_+)} }{ \sh{(\epsilon_{1,2})}  \sh(\pm 2\phi_{I} +\epsilon_{1,2}) \, \prod_{s=1}^{N} \sh{ (\pm \phi_{I} \pm a_s + \epsilon_+)} }
		\prod_{I < J}^{\ell} \frac{\sh{ ( \pm \phi_I \pm \phi_J)} \sh{ (\pm \phi_{I} \pm \phi_{J} + 2\epsilon_+) }}{\sh{ (\pm \phi_{I} \pm \phi_{J} +\epsilon_{1,2})}}\cr
  & \cdot \left( \frac{\prod_{s=1}^{N} \sh{(m \pm a_s)}}{\sh{(\pm m -\epsilon_+)}} \prod_{I=1}^{\ell} \frac{\sh{(\pm\phi_I \pm m - \epsilon_-)}}{\sh{(\pm\phi_I \pm m - \epsilon_+)}} \right)^{\chi} \prod_{I=1}^{\ell} \frac{\sh{(\pm m - \epsilon_-)} \prod_{s=1}^{N} \sh{(\pm\phi_I \pm a_s + m)} }{\sh{(\pm m - \epsilon_+)} \sh{(\pm 2\phi_I \pm m - \epsilon_+)} }\cr &\cdot \prod_{I < J}^{\ell} \frac{\sh{(\pm\phi_I \pm \phi_J \pm m - \epsilon_-)}}{\sh{(\pm\phi_I \pm \phi_J \pm m - \epsilon_+)}}\nonumber
\end{align}
\begin{equation}
  \begin{split}
&\cI_{k=2\ell+1}^-\\=&\frac{1}{\sh{ (\epsilon_{1,2})} \, \prod_{s=1}^{N} \ch{ (\pm a_s + \epsilon_+ )}} \cdot \prod_{I=1}^{\ell} \frac{ \ch{(\pm \phi_I)} \ch{ (\pm \phi_I + 2\epsilon_+)}} {\ch{(\pm \phi_I +\epsilon_{1,2} )}}\\
	    &\cdot \prod_{I=1}^{\ell} \frac{\sh{(2\epsilon_+)} }{ \sh{(\epsilon_{1,2})}  \sh{ (\pm 2\phi_{I} +\epsilon_{1,2})} \, \prod_{s=1}^{N} \sh{ (\pm \phi_{I} \pm a_s + \epsilon_+)} }
		\prod_{I < J}^{\ell} \frac{ \sh{ ( \pm \phi_I \pm \phi_J)} \sh{(\pm \phi_{I} \pm \phi_{J} + 2\epsilon_+) }}{\sh{ (\pm \phi_{I} \pm \phi_{J} +\epsilon_{1,2})}}\\
  &\cdot  \frac{\prod_{s=1}^{N} \ch{(m \pm a_s)}}{\sh{(\pm m -\epsilon_+)}} \cdot \prod_{I=1}^{\ell} \frac{\ch{(\pm\phi_I \pm m - \epsilon_-)}}{\ch{(\pm\phi_I \pm m - \epsilon_+)}}  \frac{\sh{(\pm m - \epsilon_-)} \prod_{s=1}^{N} \sh{(\pm\phi_I \pm a_s + m)} }{\sh{(\pm m - \epsilon_+)} \sh{(\pm 2\phi_I \pm m - \epsilon_+)} } \cr
  &\cdot \prod_{I < J}^{\ell} \frac{\sh{(\pm\phi_I \pm \phi_J \pm m - \epsilon_-)}}{\sh{(\pm\phi_I \pm \phi_J \pm m - \epsilon_+)}}
\end{split}
\end{equation}
\begin{align}
&\cI_{k=2\ell}^-\cr=&\frac{\ch{(\pm m - \epsilon_-)} \prod_{s=1}^{N} \sh{(2m \pm 2a_s)}}{\sh{(\pm m -\epsilon_+)}\, \sh{(\pm 2m-2\epsilon_+)}} \\
	    &\cdot \frac{\ch{(2\epsilon_+)}}{\sh{ (\epsilon_{1,2})} \,\sh{ (2\epsilon_{1,2})} \, \prod_{s=1}^{N} \sh{ (\pm 2a_s + 2\epsilon_+)}} \cdot \prod_{I=1}^{\ell-1} \frac{ \sh{(\pm 2\phi_I)}\sh{ (\pm 2\phi_I + 4\epsilon_+) } } {\sh{ (\pm 2\phi_I +2\epsilon_{1,2})} } \cr
	    &\cdot \prod_{I=1}^{\ell-1} \frac{\sh{(2\epsilon_+)} }{ \sh{(\epsilon_{1,2} )}  \sh{ (\pm 2\phi_{I} +\epsilon_{1,2})} \, \prod_{s=1}^{N} \sh{ (\pm \phi_{I} \pm a_s + \epsilon_+)} }
		\prod_{I < J}^{\ell-1} \frac{ \sh{ ( \pm \phi_I \pm \phi_J)}\sh{ (\pm \phi_{I} \pm \phi_{J} + 2\epsilon_+) }}{\sh{ (\pm \phi_{I} \pm \phi_{J} +\epsilon_{1,2})}} \cr
  &\cdot \prod_{I=1}^{\ell-1} \frac{\sh{(\pm2\phi_I \pm 2m - 2\epsilon_-)}}{\sh{(\pm2\phi_I \pm 2m - 2\epsilon_+)}} \frac{\sh{(\pm m - \epsilon_-)} \prod_{s=1}^{N} \sh{(\pm\phi_I \pm a_s + m)} }{\sh{(\pm m - \epsilon_+)} \sh{(\pm 2\phi_I \pm m - \epsilon_+)} } \cdot
    \prod_{I<J}^{\ell-1} \frac{\sh{(\pm\phi_I \pm \phi_J \pm m - \epsilon_-)}}{\sh{(\pm\phi_I \pm \phi_J \pm m - \epsilon_+)}}\nonumber
\end{align}
\end{footnotesize}

\section{Multiplicity coefficients}\label{app:coefficient}

In this Appendix, we list some of the multiplicity coefficients $C^{\textrm{4d}}_{\vec{\lambda},\vec{A}}$ in \eqref{4d-YT}. The additional effective Coulomb branch parameters that appear in \eqref{SO-adj} and \eqref{Sp-even-plus}--\eqref{Sp-odd-minus} are 
\be  
a_{N+j}=  \frac{\hbar}{2} (+\pi i )~, \   \frac{m}{2} (+\pi i )~, \   \frac{\hbar-m}{2} (+\pi i )~, \ 0 (+\pi i )~, \ \hbar (+\pi i )~, \ m (+\pi i )~.
\ee
In order to express the multiplicity coefficients $C^{\textrm{4d}}_{\lambda,A}$ that appear in these parameters, we need to introduce some notation for 4d Young diagrams. As in (\ref{4dphi(s)}), the $(x_1,x_2)$ direction is associated with the equivariant parameter $\hbar$, while the $(x_3,x_4)$ direction is associated with the equivariant parameter $m$. We represent an $(x_1,x_2)$ Young sub-diagram by drawing it explicitly, and we assign an $(x_3,x_4)$ Young sub-diagram to each box of the $(x_1,x_2)$ Young sub-diagram by monotonically decreasing positive integers $\mu=(\mu_1,\mu_2,\ldots)$. Note that we write the transposition of a 2d Young diagram $\mu$ by $\mu^t$. For example, all the contents $x=(x_1,x_2,x_3,x_4)\in \lambda$  of the following presentation of a 4d Young diagram $\lambda$ are given by
\begin{align}
\raisebox{-0.25cm}{\includegraphics[width=1.6cm]{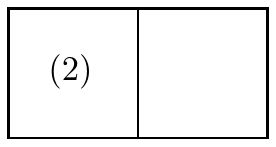}} \leadsto\quad&\{(1,1,1,1), (1,2,1,1) , (1,1,1,2)\}\cr 
      \raisebox{-0.25cm}{\includegraphics[width=1.6cm]{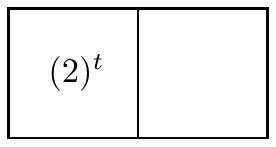}}  \leadsto\quad&\{(1,1,1,1), (1,2,1,1) , (1,1,2,1)\}\cr  
    \raisebox{-0.25cm}{\includegraphics[width=.8cm]{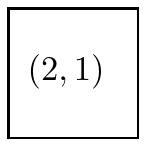}}\leadsto \quad&\{(1,1,1,1), (1,1,1,2) , (1,1,2,1)\}
\end{align}
Although we obtain multiplicity coefficients for 4d Young diagrams with total 4 boxes, they are quite complicated. Therefore, we provide a list of multiplicity coefficients for 4d Young diagrams with up to total 3 boxes that correspond to non-trivial residues.  Note that residues are trivial upto 7-instanton for 4d Young diagrams that are not listed here. For 4 boxes, the reader can refer to the Mathematica file provided on the arXiv page. 

\bigskip

\noindent $\bullet$ $C^{\textrm{4d}}_{\lambda,A=\pm q^{\frac12}} \quad (a=\frac\hbar2(+\pi i)$):
\begin{align*}
&\raisebox{-0.25cm}{\includegraphics[width=.8cm]{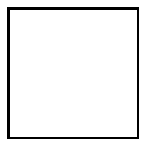}}\leadsto2;
\qquad\raisebox{-0.6cm}{\includegraphics[width=.8cm]{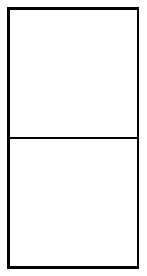}}\leadsto2;\cr
&\raisebox{-0.25cm}{\includegraphics[width=.8cm]{pictures/4dYT07}}\leadsto2;
\qquad\raisebox{-0.25cm}{\includegraphics[width=2.1cm]{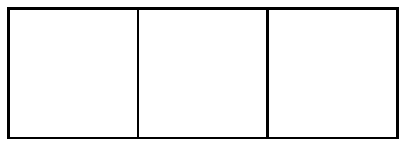}}\leadsto2; 
\qquad\raisebox{-0.55cm}{\includegraphics[width=1.4cm]{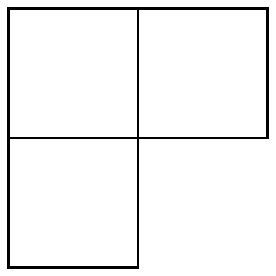}}\leadsto2;
\qquad\raisebox{-.9cm}{\includegraphics[width=.8cm]{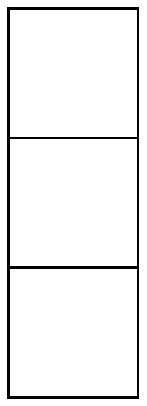}}\leadsto2;
\end{align*}

\noindent $\bullet$ $C^{\textrm{4d}}_{\lambda,A=\pm M^{\frac12}} \quad  (a=\frac{m}2(+\pi i)$):
\begin{align*}
  &  \raisebox{-0.25cm}{\includegraphics[width=.8cm]{pictures/4dYT01}}\leadsto-2;
    \qquad\raisebox{-0.25cm}{\includegraphics[width=.8cm]{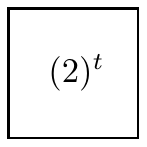}}\leadsto2;\cr 
    &\raisebox{-0.55cm}{\includegraphics[width=1.4cm]{pictures/4dYT14}}\leadsto-2;
    \qquad\raisebox{-0.25cm}{\includegraphics[width=.8cm]{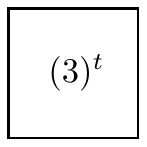}}\leadsto-2;
       \qquad\raisebox{-0.25cm}{\includegraphics[width=.8cm]{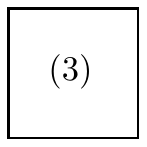}}\leadsto-2;
    \qquad\raisebox{-0.25cm}{\includegraphics[width=.8cm]{pictures/4dYT07}}\leadsto-2;
\end{align*}

\noindent $\bullet$ $C^{\textrm{4d}}_{\lambda,A=\pm \left(\frac{q}{M}\right)^{\frac12}} \quad 
 (a=\frac{\hbar-m}{2}(+i\pi)$):
\begin{align*}
    \raisebox{-0.25cm}{\includegraphics[width=1.5cm]{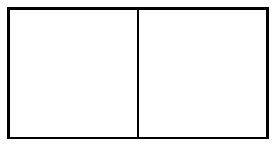}}\leadsto-1;
    \qquad\raisebox{-0.25cm}{\includegraphics[width=.8cm]{pictures/4dYT03}}\leadsto-1;
\end{align*}

\noindent $\bullet$ $C^{\textrm{4d}}_{\lambda,A=\pm 1} \quad (a=0(+i\pi))$:
\begin{align*}
 & \raisebox{-0.3cm}{\includegraphics[width=1.5cm]{pictures/4dYT04}}\leadsto2;
    \qquad\raisebox{-0.6cm}{\includegraphics[width=.8cm]{pictures/4dYT05}}\leadsto2;
    \qquad\raisebox{-0.25cm}{\includegraphics[width=.8cm]{pictures/4dYT03}}\leadsto2;
    \qquad\raisebox{-0.25cm}{\includegraphics[width=.8cm]{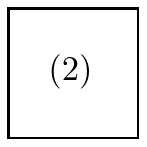}}\leadsto2; \cr
 & \raisebox{-0.25cm}{\includegraphics[width=.8cm]{pictures/4dYT10}}\leadsto-2;
    \qquad\raisebox{-0.25cm}{\includegraphics[width=.8cm]{pictures/4dYT06}}\leadsto-2;
    \qquad\raisebox{-0.25cm}{\includegraphics[width=2.1cm]{pictures/4dYT13}}\leadsto2;
    \qquad\raisebox{-1cm}{\includegraphics[width=.8cm]{pictures/4dYT15}}\leadsto2; \cr
 &\raisebox{-0.3cm}{\includegraphics[width=1.5cm]{pictures/4dYT11}}\leadsto-1;
    \qquad\raisebox{-0.25cm}{\includegraphics[width=1.5cm]{pictures/4dYT08}}\leadsto-1;
    \qquad\raisebox{-.6cm}{\includegraphics[width=.8cm]{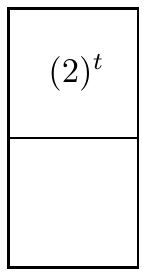}}\leadsto-1;
    \qquad\raisebox{-.6cm}{\includegraphics[width=.8cm]{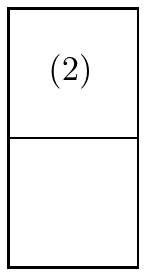}}\leadsto-1; 
\end{align*}

\noindent $\bullet$ $C^{\textrm{4d}}_{\lambda,A=\pm q} \quad (a=\hbar(+i\pi))$:
\begin{align*}
   & \raisebox{-0.25cm}{\includegraphics[width=.8cm]{pictures/4dYT01}}\leadsto1;
    \qquad\raisebox{-0.6cm}{\includegraphics[width=.8cm]{pictures/4dYT05}}\leadsto1;
    \qquad\raisebox{-0.3cm}{\includegraphics[width=1.5cm]{pictures/4dYT04}}\leadsto2;\cr 
  &\raisebox{-1cm}{\includegraphics[width=.8cm]{pictures/4dYT15}}\leadsto1;
   \qquad  \raisebox{-0.25cm}{\includegraphics[width=.8cm]{pictures/4dYT07}}\leadsto2;
    \qquad\raisebox{-0.55cm}{\includegraphics[width=1.4cm]{pictures/4dYT14}}\leadsto2;
\end{align*}

\noindent $\bullet$ $C^{\textrm{4d}}_{\lambda,A=\pm M} \quad (a=m(+i\pi))$:
\begin{align*}
    \raisebox{-0.25cm}{\includegraphics[width=.8cm]{pictures/4dYT01}}\leadsto-1;
    \qquad\raisebox{-0.25cm}{\includegraphics[width=.8cm]{pictures/4dYT03}}\leadsto1;
    \qquad\raisebox{-0.25cm}{\includegraphics[width=.8cm]{pictures/4dYT02}}\leadsto-2;
    \qquad\raisebox{-0.25cm}{\includegraphics[width=.8cm]{pictures/4dYT10}}\leadsto-1; 
    \qquad  \raisebox{-0.25cm}{\includegraphics[width=.8cm]{pictures/4dYT07}}\leadsto2;
\end{align*}

\section{Gopakumar-Vafa invariants for \texorpdfstring{$D$}{D}-type singularities}
In this appendix, we will list the Gopakumar-Vafa (GV) invariants \cite{Gopakumar:1998ii,Gopakumar:1998jq} from $\SO(2N)$ instanton partition functions.

Compactification of M-theory on a Calabi-Yau manifold leads to a five-dimensional (5d) supersymmetric theory with eight supercharges, known as \emph{geometric engineering} \cite{Katz:1996fh,Katz:1997eq}. In particular, when a Calabi-Yau manifold is a bundle in which an ALE space (a blowup of an $ADE$-orbifold singularity $\bC^2/\Gamma$) is fibered over $\boldsymbol{P}^1$, it gives rise to 5d pure Yang-Mills theory for a gauge group of $ADE$ type. In this construction, $W$-bosons arise as M2-branes wrapped on two-cycles in the ALE space, while instantons appear by wrapping M2-branes on the base $\boldsymbol{P}^1$.

Through geometric engineering, instantons are related to the Gopakumar-Vafa (GV) invariants \cite{Gopakumar:1998ii,Gopakumar:1998jq} for a Calabi-Yau manifold. These invariants are graded dimensions of BPS states that arise from M2-branes wrapping two-cycles of a Calabi-Yau threefold in M-theory, and they can be defined using techniques in algebraic geometry \cite{pandharipande2010stable,Maulik:2016rip}. As a result, the GV invariants are a set of integers, and the closed topological string partition function for a Calabi-Yau threefold becomes the GV generating function. Therefore, geometric engineering relates the GV invariants to instantons in the 5d supersymmetric theory.

 The relationship between the closed topological string partition function and the instanton partition functions has been studied extensively in the case of gauge groups of $A$ type \cite{Iqbal:2003ix,Iqbal:2003zz,Eguchi:2003sj}. Since the $\SO(2N)$ instanton partition function has been obtained in \cite{Nawata:2021dlk}, we can read off GV invariants of the ALE space with $D$-type singularity fibered over $\boldsymbol{P}^1$ from it.

To understand this relationship, we first recall the GV expansion of the free energy of a closed topological string amplitude, which is given by spins as follows:
\begin{equation}
F=\sum_{C \in H_{2}(X, \mathbb{Z})} \sum_{d=1}^{\infty} \sum_{j_{L}}(-1)^{2 j_{L}} n_{C}^{j_{L}} e^{-k T_{C}}\left(\frac{q^{-2 j_{L} d}+\cdots+q^{+2 j_{L} d}}{d\left(q^{d / 2}-q^{-d / 2}\right)^{2}}\right),\quad q=e^{i g_{s}}.
\end{equation}
Here, $n_C\in\bZ$ are invariants associated to the two-cycle $C\in H_{2}(X, \mathbb{Z})$ of a Calabi-Yau manifold $X$.
This partition function has the following infinite product form:
\begin{equation}
\begin{aligned}
Z\left(\omega, g_{s}\right) &=\exp (F)\
&=\prod_{C \in H_{2}(X, \mathbb{Z})} \prod_{j_{L}} \prod_{p=-j_{L}}^{+j_{L}} \prod_{m=0}^{\infty}\left(1-q^{2 p+m+1} Q^{C}\right)^{(-1)^{2 j_{L}+1}(m+1) n_{C}^{j_{L}}}~,
\end{aligned}
\end{equation}
where $Q^C=e^{-\int_{C} \omega}$ is the K\"{a}hler parameter associated to the holomorphic curve $C$ in $H_2(X,\mathbb{Z})$.

\subsection{\texorpdfstring{$\SO(4)$}{SO(4)}}
The K\"{a}hler parameters for $\SO(4)$ instanton partition function are $Q$ and $Q_{1,2}$. Their relations with the Coulomb branch parameters are
\begin{equation}
    A_{s}:=\prod_{j=1}^s Q_j.
\end{equation}
We can choose such a basis that $Q=\mathfrak{q}$ and $k,a,b \in \mathbb{Z}_{+}$, then $k$ is identified with instanton number
\begin{equation}
Z_{\rm inst.}=\prod_{k,a,b\in \mathbb{Z}_{+}} \prod_{j_{L}} \prod_{p=-j_{L}}^{+j_{L}} \prod_{m=0}^{\infty}\left(1-q^{2 p+m+1} Q^k Q_1^a Q_2^b\right)^{(-1)^{2 j_{L}+1}(m+1) n_{k,a,b}^{j_{L}}}
\end{equation}

The GV invariants at $k=1$ can be expressed as follows:
\begin{equation}
\begin{aligned}
n^{0}_{1,0,0} &=1,\\
n^{0}_{1,0,p} &= n^{0}_{1,2p,p} = 2p.
\end{aligned}
\end{equation}
For the $k=2$ case, we have read off the GV invariants from the instanton partition function up to $Q_1^{14}$ and $Q_2^7$. These results are summarized in Table \ref{GVso4k2}.

\begin{table}
  \centering
\begin{tabular}{cccccccccc}
\toprule
$j_L$&$a,b$&$n$&$a,b$&$n$&$j_L$&$a,b$&$n$&$a,b$&$n$ \\
\midrule
2&0,7&18&14,7&18&\multirowcell{2}{1/2}&0,6&-46&12,6&-46\\
\cmidrule{1-5}
\multirowcell{2}{3/2}&0,6&-15&12,6&-15&&0,7&-76&14,7&-76\\
\cmidrule(lr){6-10}
&0,7&-32&14,7&-32&\multirowcell{5}{0}&0,3&6&6,3&6\\
\cmidrule(lr){1-5}
\multirowcell{3}{1}&0,5&12&10,5&12&&0,4&14&8,4&14\\
&0,6&26&12,6&26&&0,5&34&10,5&34\\
&0,7&58&14,7&58&&0,6&58&12,6&58\\
\cmidrule(lr){1-5}
1/2&0,5&-20&10,5&-20&&0,7&100&14,7&100\\
\bottomrule
\end{tabular}
\caption{\label{GVso4k2}GV invariants $n_{2,a,b}^{j_L}$ for $\SO(4)$, $k=2$ case.}
\end{table}
From Table \ref{GVso4k2}, we can see that the GV invariants of $\SO(4)$ have a symmetry
\begin{equation}
    n^{j_L}_{k,0,p} = n^{j_L}_{k,2p,p}
\end{equation}
for $k=1,2$. We conjecture that this relation holds for arbitrary $k$.

\paragraph{Relation with $\SU(2)$}
The isomorphism $\mathfrak{so}(4)=\mathfrak{su}(2)\oplus\mathfrak{su}(2)$ leads to
\begin{equation}
\label{eq:so4su2}
Z_{\mathrm{SO}(4)}=Z_{\mathrm{SU}(2)}\left(Q_{1}\to Q_{1}Q^{1/2}_{2}\right) Z_{\mathrm{SU}(2)}\left(Q_{1}\to Q^{1/2}_{2}\right) Z_{\mathrm{U}(1)} \text {, }
\end{equation}
Recall that
\begin{equation}
    \begin{aligned}
        Z_{\SO(4)}=&\exp\Big[\sum_{d=1}^{\infty}\sum_{k,a,b\in\mathbb Z}\sum_{j_L}\frac{n^{j_L}_{k,a,b}Q^{dk}Q_1^{da}Q_2^{db}(q^{-2j_L d}+...+q^{2j_L d})}{d(q^{d/2}-q^{-d/2})^2} \Big]\\
        Z_{\SU(2)}=&\exp\Big[\sum_{d=1}^{\infty}\sum_{k,a\in\mathbb Z}\sum_{j_L}\frac{n^{j_L}_{k,a}Q^{dk}Q_1^{2d(k+a)}(q^{-2j_L d}+...+q^{2j_L d})}{d(q^{d/2}-q^{-d/2})^2} \Big]\\
        Z_{\U(1)}=&\exp\Big[\sum_{d=1}^{\infty}\sum_{k\in\mathbb Z}\sum_{j_L}\frac{n^{j_L}_{k}Q^{dk}(q^{-2j_L d}+...+q^{2j_L d})}{d(q^{d/2}-q^{-d/2})^2} \Big]
    \end{aligned}
\end{equation}
Substituting them into \eqref{eq:so4su2}, we can identify the relations between GV invariants of $\SO(4)$ and $\SU(2)$ (also $\U(1)$) as
\begin{equation}
\begin{aligned}
    n^{\SO(4)}_{j_L,k,0,p}=&n^{\SO(4)}_{j_L,k,2p,p}=n^{\SU(2)}_{j_L,k,p-k}\\n^{\SO(4)}_{j_L,k,0,0}=&n^{\U(1)}_{j_L,k}
    \end{aligned}
\end{equation}
where GV invariants of $\U(1)$ are just $n^{\U(1)}_{0,1}=1$ and all others are $0$.
\subsection{\texorpdfstring{$\SO(6)$}{SO(6)}}
The K\"{a}hler parameters are now $Q$ and $Q_{1,2,3}$.
\begin{equation}
Z^{\SO(6)}_{\rm inst.}=\prod_{k,a,b,c\in \mathbb{Z}_{+}} \prod_{j_{L}} \prod_{p=-j_{L}}^{+j_{L}} \prod_{m=0}^{\infty}\left(1-q^{2 p+m+1} Q^k Q_1^a Q_2^b Q_3^c\right)^{(-1)^{2 j_{L}+1}(m+1) n_{k,a,b,c}^{j_{L}}}.
\end{equation}

For the $k=1$ case, we computed the GV invariants up to $Q_1^6$ and $Q_{2,3}^3$. The results are summarized in Table \ref{GVso6k1}.

\begin{table}
  \centering
\begin{tabular}{ccccccccc}
\toprule
$j_L$&$a,b,c$&$n$&$a,b,c$&$n$&$a,b,c$&$n$&$a,b,c$&$n$ \\
\midrule
\multirowcell{9}{0}&0,0,0&2&2,2,1&8&6,3,1&10&0,1,3&14\\
&0,1,0&2&4,2,1&6&2,1,2&10&2,1,3&14\\
&2,1,0&2&0,0,2&6&0,2,2&12&0,2,3&18\\
&0,2,0&4&0,1,2&10&2,2,2&16&2,2,3&24\\
&4,2,0&4&0,3,0&6&4,2,2&12&4,2,3&18\\
&0,0,1&4&6,3,0&6&0,3,2&12&0,3,3&20\\
&0,1,1&6&0,3,1&10&2,3,2&18&2,3,3&30\\
&2,1,1&6&2,3,1&6&4,3,2&18&4,3,3&30\\
&0,2,1&6&4,3,1&6&6,3,2&12&6,3,3&20\\
\bottomrule
\end{tabular}
\caption{\label{GVso6k1}GV invariants $n_{1,a,b,c}^{j_L}$ for $\SO(6)$, $k=1$ case.}
\end{table}

\paragraph{Relation with $\SU(4)$}
The isomorphism $\mathfrak{so}(6)=\mathfrak{su}(4)$ leads to
\begin{equation}
    Z_{\mathrm{SO}(6)}\left(A_{1}, A_{2}, A_3\right)=Z_{\mathrm{SU}(4)}\left(A_{1}, A_{2}, A_3\right).
\end{equation}

And given
\begin{equation}
    \begin{aligned}
        Z_{\SO(6)}=&\exp\Big[\sum_{d=1}^{\infty}\sum_{k,a,b,c\in\mathbb Z}\sum_{j_L}\frac{n^{j_L}_{k,a,b,c}Q^{dk}Q_1^{da}Q_2^{db}Q_3^{dc}(q^{-2j_L d}+...+q^{2j_L d})}{d(q^{d/2}-q^{-d/2})^2} \Big]\\Z_{\SU(4)}=&\exp\Big[\sum_{d=1}^{\infty}\sum_{k,a,b,c\in\mathbb Z}\sum_{j_L}\frac{n^{j_L}_{k,a,b,c}Q^{dk}Q_1^{2da}Q_2^{d(a+c)}Q_3^{db}(q^{-2j_L d}+...+q^{2j_L d})}{d(q^{d/2}-q^{-d/2})^2} \Big]
    \end{aligned}
\end{equation}
We can conclude that by simple changes of variables, the GV invariants of $\SO(6)$ are related to those of $\SU(4)$ via
\begin{equation}
    n^{\SU(4)}_{j_L,k,a,b,c}=n^{\SO(6)}_{j_L,k,2a,a+c,b}\Longleftrightarrow n^{\SO(6)}_{j_L,k,a,b,c}=n^{\SU(4)}_{j_L,k,a/2,c,b-a/2}
\end{equation}

\subsection{\texorpdfstring{$\SO(8)$}{SO(8)}}

The first non-trivial case is the $D_4$ gauge group.
The K\"{a}hler parameters are $Q$ and $Q_{1,2,3,4}$.
\begin{equation}
Z_{\rm inst.}=\prod_{k,a,b,c,d\in \mathbb{Z}_{+}} \prod_{j_{L}} \prod_{p=-j_{L}}^{+j_{L}} \prod_{m=0}^{\infty}\left(1-q^{2 p+m+1} Q^k Q_1^a Q_2^b Q_3^c Q_4^d\right)^{(-1)^{2 j_{L}+1}(m+1) n_{k,a,b,c,d}^{j_{L}}}
\end{equation}

For the $k=1$ case, we computed the GV invariants up to $Q_1^4$ and $Q_{2,3,4}^2$. The results are summarized in Table \ref{GVso8k1}.

\begin{table}
  \centering
\begin{tabular}{ccccccccc}
\toprule
$j_L$&$a,b,c,d$&$n$&$a,b,c,d$&$n$&$a,b,c,d$&$n$&$a,b,c,d$&$n$ \\
\midrule
\multirowcell{11}{0}&0,0,0,0&2&0,0,2,0&6&2,2,1,1&10&2,1,1,2&6\\
&0,1,0,0&2&0,1,2,0&10&4,2,1,1&6&2,2,1,2&6\\
&2,1,0,0&2&2,1,2,0&10&0,0,2,1&10&0,0,2,2&12\\
&0,2,0,0&4&0,2,2,0&12&0,1,2,1&16&0,1,2,2&18\\
&4,2,0,0&4&2,2,2,0&16&2,1,2,1&16&2,1,2,2&18\\
&0,0,1,0&4&4,2,2,0&12&0,2,2,1&18&0,2,2,2&18\\
&0,1,1,0&6&0,0,0,1&2&2,2,2,1&30&4,2,2,2&18\\
&2,1,1,0&6&0,0,1,1&6&4,2,2,1&18&2,2,2,2&32\\
&0,2,1,0&6&0,1,1,1&8&0,0,0,2&4&&\\
&2,2,1,0&8&2,1,1,1&8&0,0,1,2&6&&\\
&4,2,1,0&6&0,2,1,1&6&0,1,1,2&6&&\\
\bottomrule
\end{tabular}
\caption{\label{GVso8k1}GV invariants $n_{1,a,b,c,d}^{j_L}$ for $\SO(8)$, $k=1$ case.}
\end{table}

\bibliographystyle{JHEP}
\bibliography{references}

\providecommand{\href}[2]{#2}\begingroup\raggedright\begin{thebibliography}{10}

\bibitem{Seiberg:1994aj}
N.~Seiberg and E.~Witten, \emph{{Monopoles, duality and chiral symmetry
  breaking in N=2 supersymmetric QCD}},
  \href{https://doi.org/10.1016/0550-3213(94)90214-3}{\emph{Nucl. Phys. B}
  {\bfseries 431} (1994) 484}
  [\href{https://arxiv.org/abs/hep-th/9408099}{{\ttfamily hep-th/9408099}}].

\bibitem{Seiberg:1994rs}
N.~Seiberg and E.~Witten, \emph{{Electric - magnetic duality, monopole
  condensation, and confinement in N=2 supersymmetric Yang-Mills theory}},
  \href{https://doi.org/10.1016/0550-3213(94)90124-4}{\emph{Nucl. Phys. B}
  {\bfseries 426} (1994) 19}
  [\href{https://arxiv.org/abs/hep-th/9407087}{{\ttfamily hep-th/9407087}}].

\bibitem{Nekrasov:2002qd}
N.A.~Nekrasov, \emph{{Seiberg-Witten prepotential from instanton counting}},
  \href{https://doi.org/10.4310/ATMP.2003.v7.n5.a4}{\emph{Adv. Theor. Math.
  Phys.} {\bfseries 7} (2003) 831}
  [\href{https://arxiv.org/abs/hep-th/0206161}{{\ttfamily hep-th/0206161}}].

\bibitem{AKMV}
M.~Aganagic, A.~Klemm, M.~Mari\~{n}o and C.~Vafa, \emph{{The Topological
  Vertex}}, \href{https://doi.org/10.1007/s00220-004-1162-z}{\emph{Commun.
  Math. Phys.} {\bfseries 254} (2005) 425}
  [\href{https://arxiv.org/abs/hep-th/0305132}{{\ttfamily hep-th/0305132}}].

\bibitem{Awata:2005fa}
H.~Awata and H.~Kanno, \emph{{Instanton counting, Macdonald functions and the
  moduli space of D-branes}},
  \href{https://doi.org/10.1088/1126-6708/2005/05/039}{\emph{JHEP} {\bfseries
  05} (2005) 039} [\href{https://arxiv.org/abs/hep-th/0502061}{{\ttfamily
  hep-th/0502061}}].

\bibitem{IKV}
A.~Iqbal, C.~Kozcaz and C.~Vafa, \emph{{The Refined topological vertex}},
  \href{https://doi.org/10.1088/1126-6708/2009/10/069}{\emph{JHEP} {\bfseries
  10} (2009) 069} [\href{https://arxiv.org/abs/hep-th/0701156}{{\ttfamily
  hep-th/0701156}}].

\bibitem{Alday:2009aq}
L.F.~Alday, D.~Gaiotto and Y.~Tachikawa, \emph{{Liouville Correlation Functions
  from Four-dimensional Gauge Theories}},
  \href{https://doi.org/10.1007/s11005-010-0369-5}{\emph{Lett.Math.Phys.}
  {\bfseries 91} (2010) 167} [\href{https://arxiv.org/abs/0906.3219}{{\ttfamily
  0906.3219}}].

\bibitem{Nawata:2021dlk}
S.~Nawata and R.-D.~Zhu, \emph{{Instanton counting and O-vertex}},
  \href{https://doi.org/10.1007/JHEP09(2021)190}{\emph{JHEP} {\bfseries 09}
  (2021) 190} [\href{https://arxiv.org/abs/2107.03656}{{\ttfamily
  2107.03656}}].

\bibitem{Hayashi:2020hhb}
H.~Hayashi and R.-D.~Zhu, \emph{{More on topological vertex formalism for
  5-brane webs with O5-plane}},
  \href{https://doi.org/10.1007/JHEP04(2021)292}{\emph{JHEP} {\bfseries 04}
  (2021) 292} [\href{https://arxiv.org/abs/2012.13303}{{\ttfamily
  2012.13303}}].

\bibitem{Nekrasov-Shadchin}
N.~Nekrasov and S.~Shadchin, \emph{{ABCD of instantons}},
  \href{https://doi.org/10.1007/s00220-004-1189-1}{\emph{Commun. Math. Phys.}
  {\bfseries 252} (2004) 359}
  [\href{https://arxiv.org/abs/hep-th/0404225}{{\ttfamily hep-th/0404225}}].

\bibitem{Shadchin:2005mx}
S.~Shadchin, \emph{{On certain aspects of string theory/gauge theory
  correspondence}},  {Ph.D.} thesis, 2, 2005,
  [\href{https://arxiv.org/abs/hep-th/0502180}{{\ttfamily hep-th/0502180}}].

\bibitem{Kim:2012gu}
H.-C.~Kim, S.-S.~Kim and K.~Lee, \emph{{5-dim Superconformal Index with
  Enhanced $E_n$ Global Symmetry}},
  \href{https://doi.org/10.1007/JHEP10(2012)142}{\emph{JHEP} {\bfseries 10}
  (2012) 142} [\href{https://arxiv.org/abs/1206.6781}{{\ttfamily 1206.6781}}].

\bibitem{Kim:2012qf}
H.-C.~Kim, J.~Kim and S.~Kim, \emph{{Instantons on the 5-sphere and
  M5-branes}},  \href{https://arxiv.org/abs/1211.0144}{{\ttfamily 1211.0144}}.

\bibitem{Hwang:2014uwa}
C.~Hwang, J.~Kim, S.~Kim and J.~Park, \emph{{General instanton counting and 5d
  SCFT}}, \href{https://doi.org/10.1007/JHEP07(2015)063}{\emph{JHEP} {\bfseries
  07} (2015) 063} [\href{https://arxiv.org/abs/1406.6793}{{\ttfamily
  1406.6793}}].

\bibitem{Bergman:2014kza}
O.~Bergman and G.~Zafrir, \emph{{Lifting 4d dualities to 5d}},
  \href{https://doi.org/10.1007/JHEP04(2015)141}{\emph{JHEP} {\bfseries 04}
  (2015) 141} [\href{https://arxiv.org/abs/1410.2806}{{\ttfamily 1410.2806}}].

\bibitem{Bergman:2015dpa}
O.~Bergman and G.~Zafrir, \emph{{5d fixed points from brane webs and
  O7-planes}}, \href{https://doi.org/10.1007/JHEP12(2015)163}{\emph{JHEP}
  {\bfseries 12} (2015) 163}
  [\href{https://arxiv.org/abs/1507.03860}{{\ttfamily 1507.03860}}].

\bibitem{Zafrir:2015ftn}
G.~Zafrir, \emph{{Brane webs and $O5$-planes}},
  \href{https://doi.org/10.1007/JHEP03(2016)109}{\emph{JHEP} {\bfseries 03}
  (2016) 109} [\href{https://arxiv.org/abs/1512.08114}{{\ttfamily
  1512.08114}}].

\bibitem{Hayashi:2016jak}
H.~Hayashi and G.~Zoccarato, \emph{{Partition functions of web diagrams with an
  O7$^{-}$-plane}}, \href{https://doi.org/10.1007/JHEP03(2017)112}{\emph{JHEP}
  {\bfseries 03} (2017) 112}
  [\href{https://arxiv.org/abs/1609.07381}{{\ttfamily 1609.07381}}].

\bibitem{Kim-Yagi}
S.-S.~Kim and F.~Yagi, \emph{{Topological vertex formalism with O5-plane}},
  \href{https://doi.org/10.1103/PhysRevD.97.026011}{\emph{Phys. Rev.}
  {\bfseries D97} (2018) 026011}
  [\href{https://arxiv.org/abs/1709.01928}{{\ttfamily 1709.01928}}].

\bibitem{Kim:2018gjo}
H.-C.~Kim, J.~Kim, S.~Kim, K.-H.~Lee and J.~Park, \emph{{6d strings and
  exceptional instantons}},
  \href{https://doi.org/10.1103/PhysRevD.103.025012}{\emph{Phys. Rev. D}
  {\bfseries 103} (2021) 025012}
  [\href{https://arxiv.org/abs/1801.03579}{{\ttfamily 1801.03579}}].

\bibitem{Hayashi:2018bkd}
H.~Hayashi, S.-S.~Kim, K.~Lee and F.~Yagi, \emph{{5-brane webs for 5d $
  \mathcal{N} $ = 1 G$_{2}$ gauge theories}},
  \href{https://doi.org/10.1007/JHEP03(2018)125}{\emph{JHEP} {\bfseries 03}
  (2018) 125} [\href{https://arxiv.org/abs/1801.03916}{{\ttfamily
  1801.03916}}].

\bibitem{Hayashi:2018lyv}
H.~Hayashi, S.-S.~Kim, K.~Lee and F.~Yagi, \emph{{Dualities and 5-brane webs
  for 5d rank 2 SCFTs}},
  \href{https://doi.org/10.1007/JHEP12(2018)016}{\emph{JHEP} {\bfseries 12}
  (2018) 016} [\href{https://arxiv.org/abs/1806.10569}{{\ttfamily
  1806.10569}}].

\bibitem{Cheng:2018wll}
S.~Cheng and S.-S.~Kim, \emph{{Refined topological vertex for a 5D Sp(N) gauge
  theories with antisymmetric matter}},
  \href{https://doi.org/10.1103/PhysRevD.104.086004}{\emph{Phys. Rev. D}
  {\bfseries 104} (2021) 086004}
  [\href{https://arxiv.org/abs/1809.00629}{{\ttfamily 1809.00629}}].

\bibitem{Hayashi:2019yxj}
H.~Hayashi, S.-S.~Kim, K.~Lee and F.~Yagi, \emph{{Rank-3 antisymmetric matter
  on 5-brane webs}}, \href{https://doi.org/10.1007/JHEP05(2019)133}{\emph{JHEP}
  {\bfseries 05} (2019) 133}
  [\href{https://arxiv.org/abs/1902.04754}{{\ttfamily 1902.04754}}].

\bibitem{Kim:2021cua}
H.-C.~Kim, M.~Kim and S.-S.~Kim, \emph{{Topological vertex for 6d SCFTs with
  $\mathbb{Z}_2$-twist}},
  \href{https://doi.org/10.1007/JHEP03(2021)132}{\emph{JHEP} {\bfseries 03}
  (2021) 132} [\href{https://arxiv.org/abs/2101.01030}{{\ttfamily
  2101.01030}}].

\bibitem{Kim:2022dbr}
S.-S.~Kim and X.-Y.~Wei, \emph{{Refined topological vertex with ON-planes}},
  \href{https://doi.org/10.1007/JHEP08(2022)006}{\emph{JHEP} {\bfseries 08}
  (2022) 006} [\href{https://arxiv.org/abs/2201.12264}{{\ttfamily
  2201.12264}}].

\bibitem{Losev:2003py}
A.S.~Losev, A.~Marshakov and N.A.~Nekrasov, \emph{{Small instantons, little
  strings and free fermions}},  in \emph{{From Fields to Strings:
  Circumnavigating Theoretical Physics: A Conference in Tribute to Ian Kogan}},
  pp.~581--621, 2, 2003 [\href{https://arxiv.org/abs/hep-th/0302191}{{\ttfamily
  hep-th/0302191}}].

\bibitem{jeffrey1995localization}
L.C.~Jeffrey and F.C.~Kirwan, \emph{Localization for nonabelian group actions},
  {\emph{Topology} {\bfseries 34} (1995) 291}
  [\href{https://arxiv.org/abs/alg-geom/9307001}{{\ttfamily
  alg-geom/9307001}}].

\bibitem{Benini:2013xpa}
F.~Benini, R.~Eager, K.~Hori and Y.~Tachikawa, \emph{{Elliptic Genera of 2d
  ${\mathcal{N}}$ = 2 Gauge Theories}},
  \href{https://doi.org/10.1007/s00220-014-2210-y}{\emph{Commun. Math. Phys.}
  {\bfseries 333} (2015) 1241}
  [\href{https://arxiv.org/abs/1308.4896}{{\ttfamily 1308.4896}}].

\bibitem{Chengdu}
S.-S.~Kim, X.~Li, S.~Nawata and F.~Yagi, \emph{{Higgsing and tuning D-branes
  with O-planes}}, {\emph{Work in progress} }.

\bibitem{Kim:2019uqw}
J.~Kim, S.-S.~Kim, K.-H.~Lee, K.~Lee and J.~Song, \emph{{Instantons from
  Blow-up}}, \href{https://doi.org/10.1007/JHEP11(2019)092}{\emph{JHEP}
  {\bfseries 11} (2019) 092}
  [\href{https://arxiv.org/abs/1908.11276}{{\ttfamily 1908.11276}}].

\bibitem{Sen:1996vd}
A.~Sen, \emph{{F theory and orientifolds}},
  \href{https://doi.org/10.1016/0550-3213(96)00347-1}{\emph{Nucl. Phys. B}
  {\bfseries 475} (1996) 562}
  [\href{https://arxiv.org/abs/hep-th/9605150}{{\ttfamily hep-th/9605150}}].

\bibitem{Hanany:1996ie}
A.~Hanany and E.~Witten, \emph{{Type IIB superstrings, BPS monopoles, and
  three-dimensional gauge dynamics}},
  \href{https://doi.org/10.1016/S0550-3213(97)00157-0}{\emph{Nucl. Phys. B}
  {\bfseries 492} (1997) 152}
  [\href{https://arxiv.org/abs/hep-th/9611230}{{\ttfamily hep-th/9611230}}].

\bibitem{Gopakumar:1998ii}
R.~Gopakumar and C.~Vafa, \emph{{M theory and topological strings. 1.}},
  \href{https://arxiv.org/abs/hep-th/9809187}{{\ttfamily hep-th/9809187}}.

\bibitem{Gopakumar:1998jq}
R.~Gopakumar and C.~Vafa, \emph{{M theory and topological strings. 2.}},
  \href{https://arxiv.org/abs/hep-th/9812127}{{\ttfamily hep-th/9812127}}.

\bibitem{Katz:1996fh}
S.H.~Katz, A.~Klemm and C.~Vafa, \emph{{Geometric engineering of quantum field
  theories}}, \href{https://doi.org/10.1016/S0550-3213(97)00282-4}{\emph{Nucl.
  Phys. B} {\bfseries 497} (1997) 173}
  [\href{https://arxiv.org/abs/hep-th/9609239}{{\ttfamily hep-th/9609239}}].

\bibitem{Katz:1997eq}
S.~Katz, P.~Mayr and C.~Vafa, \emph{{Mirror symmetry and exact solution of 4-D
  N=2 gauge theories: 1.}},
  \href{https://doi.org/10.4310/ATMP.1997.v1.n1.a2}{\emph{Adv. Theor. Math.
  Phys.} {\bfseries 1} (1998) 53}
  [\href{https://arxiv.org/abs/hep-th/9706110}{{\ttfamily hep-th/9706110}}].

\bibitem{pandharipande2010stable}
R.~Pandharipande and R.~Thomas, \emph{{Stable pairs and BPS invariants}},
  {\emph{Journal of the American Mathematical Society} {\bfseries 23} (2010)
  267} [\href{https://arxiv.org/abs/0711.3899}{{\ttfamily 0711.3899}}].

\bibitem{Maulik:2016rip}
D.~Maulik and Y.~Toda, \emph{{Gopakumar-Vafa invariants via vanishing cycles}},
  {\emph{Inventiones mathematicae} {\bfseries 213} (2018) 1017}
  [\href{https://arxiv.org/abs/1610.07303}{{\ttfamily 1610.07303}}].

\bibitem{Iqbal:2003ix}
A.~Iqbal and A.-K.~Kashani-Poor, \emph{{Instanton counting and Chern-Simons
  theory}}, \href{https://doi.org/10.4310/ATMP.2003.v7.n3.a4}{\emph{Adv. Theor.
  Math. Phys.} {\bfseries 7} (2003) 457}
  [\href{https://arxiv.org/abs/hep-th/0212279}{{\ttfamily hep-th/0212279}}].

\bibitem{Iqbal:2003zz}
A.~Iqbal and A.-K.~Kashani-Poor, \emph{{SU(N) geometries and topological string
  amplitudes}}, \href{https://doi.org/10.4310/ATMP.2006.v10.n1.a1}{\emph{Adv.
  Theor. Math. Phys.} {\bfseries 10} (2006) 1}
  [\href{https://arxiv.org/abs/hep-th/0306032}{{\ttfamily hep-th/0306032}}].

\bibitem{Eguchi:2003sj}
T.~Eguchi and H.~Kanno, \emph{{Topological strings and Nekrasov's formulas}},
  \href{https://doi.org/10.1088/1126-6708/2003/12/006}{\emph{JHEP} {\bfseries
  12} (2003) 006} [\href{https://arxiv.org/abs/hep-th/0310235}{{\ttfamily
  hep-th/0310235}}].

\end{thebibliography}\endgroup

\end{document}